\documentclass[12pt]{article}
\usepackage{setspace}
\usepackage{amsmath}
\usepackage{graphicx}
\usepackage{cases}
\usepackage{subfigure}
\usepackage{natbib}
\usepackage{psfrag}
\usepackage{amssymb}
\usepackage{color,graphics,epsfig}
\usepackage{appendix}
\usepackage{fancyhdr}
\usepackage{xr}
\def\be{\begin{equation}}
\def\ee{\end{equation}}
\def\bea{\begin{eqnarray}}
\def\eea{\end{eqnarray}}
\def\bml{\begin{mathletters}}
\def\eml{\end{mathletters}}

\newcommand{\tc}{\textcolor}

\begin{document}

\title{Rapid adaptation of a polygenic trait after a sudden environmental shift}

\author{Kavita Jain$^{\dagger}$ and Wolfgang Stephan$^{\S}$\\\mbox{}\\
$^{\dagger}$Theoretical
  Sciences Unit, \\Jawaharlal Nehru Centre for Advanced Scientific Research, \\Jakkur P.O., Bangalore 560064, India  \\
  $^{\S}$ Natural History Museum Berlin, Invalidenstr. 43, \\
  10115 Berlin, Germany}

\maketitle

\newpage

\noindent
Running head: Rapid adaptation
\bigskip


\bigskip

\noindent
Correspondence:
\texttt{jain@jncasr.ac.in}, \texttt{stephan@bio.lmu.de}

\bigskip
   
\noindent
\textbf{Abstract:} Although a number of studies \tc{black}{have shown that natural and laboratory populations} initially well-adapted to their environment can evolve rapidly when conditions suddenly change, the dynamics of rapid adaptation are not well understood. Here a population genetic model of polygenic selection is analyzed to describe the short-term response of a quantitative trait after a sudden shift of the phenotypic optimum. { We provide explicit analytical expressions for the time scales over which the trait mean approaches the new optimum. We find that when the effect sizes are small relative to a scaled mutation rate, the genomic signatures of polygenic selection are small to moderate allele frequency changes that occur in the short-term phase in a synergistic fashion. In contrast, selective sweeps, {\it i.e.},  dramatic changes in the allele frequency may occur provided the size of the effect is sufficiently large.}
Applications of our theoretical results to the relationship between QTL and selective sweep mapping and to tests of fast polygenic adaptation are discussed.

\newpage


In ``What Darwin got wrong", J. B. \citet{Losos:2014} persuasively argues that we can observe evolution in action and, in particular, that evolution can be so rapid that evolutionary and ecological time scales are confluent. The examples range from the peppered moth \citep{Cook:2012}, insecticide resistance in Drosophila \citep{Ffrench:2002}, color of field mice \citep{Vignieri:2010}, beak size in Darwin's finches \citep{Grant:2008}, guppies in Trinidad \citep{Reznick:2009}, and Anolis lizards \citep{Losos:2009} to name but a few. The rapid changes are responses to natural and human-induced shifts in the environment. The genetic architecture underlying these traits ranges from few genes of major effect to highly polygenic systems \citep{Hof:2011,Linnen:2013,Lamichhaney:2012,Lamichhaney:2015}. In this paper, we study a model that encompasses a wide range of genetic architectures. Our aim is to understand the genomic signatures of positive selection in these systems that occur after environmental changes with an emphasis on rapid adaptation.

	There is a growing body of literature on the detection of adaptive signatures in molecular population genetics. Following pioneering work of \citet{Smith:1974}, the impact of positive selection on neutral DNA variability (Òselective sweepsÓ) has attracted much interest. This theory has been applied to huge datasets that emerge from modern high-throughput sequencing. A large number of statistical tests have been developed to detect sweep signals and estimate the frequency and strength of selection \citep{Kim:2002,Nielsen:2005b,Pavlidis:2010}. To find sweep signatures in the genome, strong positive selection is required (with $N_e s \gg 1$, where $N_e$ is the effective population size and $s$ the selective advantage of a beneficial allele) \citep{Kaplan:1989,Stephan:1992}. Thus, sweeps are characteristic signals of fast adaptation. They are expected to be found at individual genes or at major loci if a trait is controlled by multiple genes. 
		
This raises the question whether fast evolution is also possible in highly polygenic systems and, if so, which genomic signatures arise. 
A number of genome-wide association studies (GWAS) have shown that quantitative traits are typically highly polygenic (e.g., \citet{Turchin:2012}) and there is growing evidence that the molecular scenario of sweeps only covers part of the adaptive process and needs to be revised to include polygenic selection \citep{Pritchard:2010}. As GWAS yield information about the distribution of single nucleotide polymorphisms (SNPs) relevant to quantitative traits \citep{Visscher:2012}, it is important to understand the models of polygenic selection in terms of the frequency changes of molecular variants, {\it i.e.}, with reference to population genetics \citep{Burger:2000}.  

Although the equilibrium structure of the allele frequencies and the stationary variance have been the subject of research in a large number of such studies \citep{Burger:2000}, the dynamics have received relatively little attention.  
Perhaps the most well studied model in this context is Lande's model \citep{Lande:1983} in which the changes in the allele frequency at a major locus in the background of a large number of minor loci are considered \citep{Chevin:2008,Gomulkiewicz:2010}. However in Lande's model, the background is not explicitly modeled and it is assumed that the background variance does not evolve. \citet{Pavlidis:2012} have studied a more detailed model but their analysis is restricted to a very small number of loci. 

Following our preliminary study \citep{Jain:2015}, here we perform a population genetic analysis of the dynamics of a  polygenic trait after a sudden environmental shift of the phenotypic optimum. We consider a quantitative trait that is determined additively by a large number of diallelic loci of unequal effects. The population is assumed to be infinitely large and to evolve under stabilizing selection and mutation. This model was first proposed by \citet{Wright:1935} and more recently revisited by N. H. Barton \citep{Barton:1986,Vladar:2014}. We carry out a detailed study of this model to understand the dynamics of allele frequencies as well as of the trait mean and variance. 
In particular, to describe rapid adaptive evolution, we concentrate on the short-term period after the environmental change, which may be defined as the time until the phenotypic mean reaches a value close to the new optimum. 

{We find that the short-time dynamics are well described by a {\it directional selection model} which is analytically tractable. Working within the framework of this simpler model, 
  we reproduce some results obtained using different models or assumptions \citep{Lande:1983,Chevin:2008,Jain:2015} when most effects are small relative to a scaled mutation rate \citep{Vladar:2014}. In addition, we obtain several new results, in particular when most effects are large.}


\bigskip
\centerline{MODEL}
\bigskip

We consider an infinitely large population of diploids in which a trait $z$ is determined by $\ell$ diallelic loci. At the $i$th locus, let the $+$ allele have an effect $\gamma_i/2$ and frequency $p_i$ and, correspondingly, the $-$ allele have an effect $-\gamma_i/2$ and frequency $q_i=1-p_i$ \citep{Vladar:2014,Jain:2015}. The effect at each locus is assumed to be  exponentially distributed with mean ${\bar \gamma}$, as is often the case in quantitative genetic studies \citep{Mackay:2004,Goddard:2009}. 
Then, on neglecting dominance and epistasis, the \tc{black}{trait $z$ is} determined additively and its mean $c_1$ can be written as 
\be
c_1= \sum_{i=1}^\ell (\gamma_i p_i -\gamma_i q_i) = \sum_{i=1}^\ell \gamma_i (2 p_i- 1)~.
\label{meandefn}
\ee

For a phenotypic trait under stabilizing selection, assuming that the deviation of the fitness from its optimum $z'$ is quadratic \citep{Burger:2000}, we can write the phenotypic fitness as $w(z)=1- (s/2) (z-z')^2$, where \tc{black}{$s$ measures the strength of stabilizing selection on the trait.}  \tc{black}{Averaging over the phenotypic trait distribution (which is not necessarily Gaussian), one finds the average fitness to be}
\tc{black}
{\bea
{\overline w} &=& 1- (s/2) \overline{(z^2-c_1^2)+(c_1^2-2 z z' +z^{'2})} ~, \\
&\approx& \exp \left[-(s/2) (c_2+(c_1-z')^2) \right] ~,
\label{wbardefn}
\eea} 
where $c_2$ is the variance of the trait and is given by (see Appendix~\ref{app_cum})
\be
c_2 = 2 \sum_{i=1}^\ell \gamma_i^2 p_i q_i \label{vardefn}~. 
\ee

When selection is weak and recombination rate is high (linkage equilibrium; {see also Appendix~\ref{app_ld}}) and mutations are absent, the allele frequency $p_i$ evolves according to Wright's equation as (Chapter 6, \citet{Burger:2000})
\be
{\dot p}_i = \frac{p_i (1-p_i)}{2{\overline w}} \frac{\partial {\overline w}}{\partial p_i}~,
\label{Wright}
\ee
where dot denotes the derivative with respect to time. For the model described above, we then have  \citep{Vladar:2014,Jain:2015}
\bea
{\dot p}_i= -s \gamma_i p_i q_i  (c_1-z')-\frac{s \gamma_i^2}{2} p_i q_i(q_i-p_i) +\mu (q_i-p_i)
~,~i=1,...,\ell ~.
\label{pevoleqnf}  
\eea
In the above equation, the first two terms on the RHS follow \tc{black}{from (\ref{wbardefn})} and (\ref{Wright}). The first term tends to stabilize the phenotypic mean to the optimum trait value and the second term results in the fixation of one of the alleles thus depleting genetic variation. {The last term on the RHS \tc{black}{models the mutation process restoring} genetic variance \citep{Barton:1986}, and describes changes between the $+$ and $-$ allele at locus $i$ with locus-independent, symmetric rate $\mu$.}
In the following, we refer to the model defined by (\ref{pevoleqnf}) as the {\it full model}. 

In this article, we are interested in the response of the phenotypic mean and variance  when the phenotypic optimum is suddenly shifted to a new value $z_f$ in a population initially equilibrated to $z_o=0$. A preliminary investigation of such a situation has been carried out numerically by \citet{Vladar:2014}, and our goal here is to provide an analytical understanding of the dynamics. We will  thus focus on the dynamics of the  allele frequency that evolves according to (\ref{pevoleqnf}) when the phenotypic optimum $z'=z_f$, starting from the stationary state frequency $p_i^*$ of the population that is equilibrated to a phenotypic optimum at zero. 

Unless stated otherwise, in this article, we assume that $z_f > 0$ so that the initial mean deviation from the phenotypic optimum, $\Delta c_1(0)=-z_f$, is negative. {We also restrict the magnitude of the shift such that $z_f < \ell {\bar \gamma}$; this is because the first term on the RHS of (\ref{pevoleqnf}) acts to minimize the deviation of the trait mean from its optimum. But, from (\ref{meandefn}), the magnitude of the mean can not exceed $\sum_{i=1}^\ell \gamma_i \approx \ell {\bar \gamma}$ and therefore a shift beyond the total effect of all loci is not within evolutionary reach.}

\bigskip
\centerline{STATIONARY STATE OF THE FULL MODEL}
\bigskip

{Before proceeding further, we briefly review the known results for the stationary state that are pertinent to the later discussion.
In the stationary state where the allele frequencies are independent of time, setting the LHS of (\ref{pevoleqnf}) to zero and performing some simple algebra, we find that the equilibrium allele frequency $p_i^{*}$ is a solution of the following equation,}
\be
p_i^{*3}-p_i^{*2} \left(\frac{3}{2}+\frac{\Delta c_1^*}{\gamma_i} \right)+p_i^{*} \left[\frac{1}{2}+\frac{\Delta c_1^*}{\gamma_i} +\frac{1}{4} \left(\frac{{\hat \gamma}}{\gamma_i} \right)^2 \right] -\frac{1}{8} \left(\frac{{\hat \gamma}}{\gamma_i} \right)^2 =0~,
\label{sseqn}
\ee
where $\hat \gamma=2 \sqrt{2 \mu/s}$ and $\Delta c_1^*$ is the deviation of the mean from the optimum in equilibrium.  

When the stationary mean deviation $\Delta c_1^*$ is zero, equation (\ref{sseqn}) for the equilibrium allele frequency has three solutions that are given by \citep{Vladar:2014}
\be
{p_i^*= }
\begin{cases}
\frac{1}{2} ~&,~ \gamma_i < {\hat \gamma} \\
\frac{1}{2} \pm \frac{1}{2} \sqrt{1- \frac{\hat \gamma^2}{\gamma_i^2}} ~&,~ \gamma_i > {\hat \gamma}~.
\end{cases}
\label{icss}
\ee
While the solution $p_i^*=1/2$ is stable when the effect is smaller than the threshold  effect $\hat \gamma$, the two latter ones hold when the effect is larger than $\hat \gamma$. \tc{black}{The threshold $\hat \gamma$ between large- and small-effect alleles arises because of mutation. When selection is much weaker than mutation, the equilibrium frequency is one half; {\it i.e.},  when the stationary trait mean is at the optimum, mutation and selection balance each other at an intermediate allele frequency. The value of one half arises because we assumed that mutation in both directions is symmetric (see above). Relaxing this symmetry assumption would lead to different equilibrium frequencies. However, since we focus in the following on the short-term behavior of the dynamics, detailed mutation models are of relatively little importance compared to selection and will not be considered here.} For exponentially distributed effects, as the fraction of loci with large effects is  given by $\bar \gamma^{-1}\int_{{\hat \gamma}}^\infty dx ~   e^{-{x}/{\bar \gamma}} 
=e^{-{\hat \gamma}/{\bar \gamma}}$, 
{most effects are large} when ${\hat \gamma} \ll {\bar \gamma}$, while {most effects are small} in the opposite parameter regime. 

Furthermore, on separating the contribution from loci with small and large effects and using the stationary state frequency (\ref{icss}), we obtain the stationary state variance to be \citep{Vladar:2014,Jain:2015}
\be
c_2^*={\ell {\bar \gamma}^2}  \left[1-\left(1+ \frac{\hat \gamma}{\bar \gamma} \right) e^{-{\hat \gamma}/{\bar \gamma}} \right]~.
\label{ssvar}
\ee
The above result shows that when most effects are small,  $c_2^* \approx \ell {\bar \gamma}^2$ while in the opposite case, it is well approximated by $\ell {\hat \gamma}^2$. 

The correction to the equilibrium allele frequency in (\ref{icss}) when the stationary mean deviation is nonzero but small can be found in the Appendix B of \citet{Vladar:2014}. For future reference, we note that these corrections are negligible except when the effect is close to the threshold ${\hat \gamma}$ (see Fig.~\ref{fig_small_freq} below). 

\bigskip
\centerline{DYNAMICS WHEN MOST EFFECTS ARE SMALL}
\bigskip

We now study the dynamics of the allele frequency when most effects are small (${\bar \gamma} \ll {\hat \gamma}$). 
Since the population is initially equilibrated to the phenotypic optimum at zero, the initial frequency at the $i$th locus is well approximated by (\ref{icss}) and close to one half when most effects are small. Then, as $p_i(t) \approx q_i(t)$ at short times, we can neglect the last two terms on the RHS of (\ref{pevoleqnf}) compared to the first term to get \citep{Jain:2015}
{\be
{\dot p_i} = S_i p_i q_i  ~,
\label{pevolshort}
\ee
where $S_i=-s \gamma_i \Delta c_1$ and $\Delta c_1=c_1-z_f$. 
When the final mean deviation is negligible, the final allele frequency is also close to one half and we expect that (\ref{pevolshort}) can describe the bulk of the allele frequency dynamics as is indeed supported by Figs.~\ref{fig_small_mean1} and \ref{fig_small_freq}.}


\noindent{\bf Directional selection model:} In the following, we will refer to the model defined by (\ref{pevolshort}) as the {\it directional selection model}. 
Appendix~\ref{app_genl} details the prescription for calculating relevant quantities analytically in this model. 
{As shown in Appendix~\ref{app_genl}, the trait mean can be written as
\be
c_1(t)= \sum_{i=1}^\ell \gamma_i -\sum_{i=1}^\ell \frac{2 \gamma_i}{1+\frac{p_i(0)}{q_i(0)} ~e^{\beta {\gamma_i}/{\bar \gamma}}}~,
\label{app_c1p1}
\ee
where the parameter $\beta$ defined in (\ref{app_omegadefn}) is proportional to the logarithm of the ratio of allele frequencies and evolves in time as 
\be
{\dot \beta}=-s {\bar \gamma} \Delta c_1. 
\label{betaeqn}
\ee
Thus a closed equation for $\beta(t)$ can be obtained by eliminating the mean using (\ref{app_c1p1}) and on plugging the result for $\beta(t)$ back in (\ref{app_c1p1}), the mean can be found. 
Furthermore, it is shown in Appendix~\ref{app_genl} that in the directional selection model, the $n$th cumulant of the effect evolves according to \citep{Burger:1991}
\be
{\dot c}_n  = -s  \Delta c_1 c_{n+1} ~,~n \geq 1 ~,
\label{cumshort}
\ee
which shows that the $n$th cumulant $c_n, n \geq 2, $ can be found once $c_{n-1}$ is known. 
Thus to describe the short-time dynamics, we will focus on the key equations (\ref{app_c1p1}) and (\ref{betaeqn}).}

\tc{black}{We now return to (\ref{app_c1p1}) for the mean $c_1$ and analyze it when the number of loci is large.} 
When most effects are small (${\hat \gamma \gg \bar \gamma}$), as mentioned above, the initial allele frequency is one half. {Furthermore, as explained in DISCUSSION, the sum over the contribution from individual loci in (\ref{app_c1p1}) amounts to averaging over the distribution of effects when the number of loci is large.} 
For large $\ell$, on approximating the sum  in (\ref{app_c1p1}) by an integral,  we thus get 
\tc{black}{\be
c_1(t) =  {\ell {\bar \gamma} - \frac{\ell}{\bar \gamma} \int_0^{\hat \gamma} d\gamma \frac{ 2 \gamma e^{-\gamma/{\bar \gamma}}}{1+e^{\beta {\gamma}/{\bar \gamma}}}} ~.
\ee
Writing $x=\gamma/{\bar \gamma}$ in the above integral, we get 
\be
c_1(t)= \ell {\bar \gamma} (1 -2  I_s(\beta)) ~,
\label{small_mean2}
\ee
where}
\be
I_s(\beta) = \int_0^{{\hat \gamma}/{\bar \gamma}} dx ~\frac{x e^{-x}}{1+e^{\beta x}} ~.
\ee
Our main task is to find the time dependence of $\beta$ using (\ref{betaeqn}) and (\ref{small_mean2}) which is dealt with in Appendix~\ref{app_small}. 


\noindent{\bf {Qualitative patterns of the cumulant dynamics:}} 
Before turning to explicit expressions, we first make some general remarks on the behavior of the cumulants in the  directional selection model. Since the variance is always nonnegative, it follows from (\ref{cumshort}) that the magnitude of mean deviation decreases with time. As we are considering the scenario where $\Delta c_1(0) < 0$, the mean deviation $\Delta c_1(t)$ increases {\it monotonically} with time towards zero as shown in Fig.~\ref{fig_small_mean1}. 
Due to (\ref{betaeqn}), this also means that $\beta(t)$ initially increases and then saturates to a constant.

Furthermore, due to (\ref{betaeqn}) and (\ref{cumshort}), the variance can be written as 
\be
c_2=-2 \ell {\bar \gamma}^2 I_s'(\beta) ~,
\label{c2Is}
\ee
where prime denotes the derivative with respect to $\beta$. \tc{black}{To arrive at the above expression, we have used that ${\dot c}_1=-2 \ell {\bar \gamma} I_s'(\beta) {\dot \beta}$ due to (\ref{small_mean2})}.  Equation (\ref{c2Is}) gives the rate of change of variance as 
${\dot c}_2= -2 \ell {\bar \gamma}^2 {\dot \beta} ~I_s''(\beta)$. It is easy to check that the integrand of $I_s''$ is always positive from which it follows that $I_s''(\beta) > 0$. Therefore the directional selection model predicts that when most effects are small, the variance {\it decreases} with time as indeed verified in Fig.~\ref{fig_small_mean1}. From this result and (\ref{cumshort}), it  immediately follows that the skewness $c_3$ defined in (\ref{app_skewdefn}) is always negative. 

\noindent{\bf {Quantitative dynamics of the mean and higher cumulants:}}
As shown in Appendix~\ref{app_small}, the mean deviation vanishes  exponentially fast,
\be
\Delta c_1(t)= \Delta c_1(0) e^{-s c_2(0) t}~,
\label{small_mean}
\ee
where $c_2(0) \approx \ell {\bar \gamma}^2$ is the initial variance given by (\ref{ssvar}) when most effects are small. {The above result has been previously obtained by assuming that the variance and higher cumulants do not evolve in time \citep{Lande:1983,Chevin:2008,Jain:2015}. The basis of this assumption lies in phenotypic data \citep{Lande:1976} or is motivated by mathematical tractability \citep{Chevin:2008,Gomulkiewicz:2010,Jain:2015}. Here, in contrast, we have obtained (\ref{small_mean}) without any additional assumptions.}

Using the solution (\ref{small_mean}) in (\ref{cumshort}) \tc{black}{for cumulant dynamics}, we find that the variance stays at its initial value $c_2(0)$ and the higher cumulants vanish. {The corrections to these behavior at late times are given by (\ref{app_mvsma1}) and (\ref{app_mvsma2}), and we find that the variance remains a constant at short times ($\ll (s c_2(0))^{-1}$) but decreases thereafter to a smaller value:
\be
\frac{c_2(t)}{c_2(0)} \approx 1 - 3 \left(\frac{z_f}{\ell {\bar \gamma}}\right)^2 (1-e^{-s c_2(0) t})^2~.
\label{app_varsma}
\ee
The above expression shows that the drop in the variance is larger for large shifts in the optimum and can be quite significant as shown in Fig.~\ref{fig_small_mean1} for a moderate shift (relative to the maximum mean).} 
A comparison between the full model, directional selection model and the analytical results is also shown in Fig.~\ref{fig_small_mean1} for mean and variance. The reason for the difference observed in Fig.~\ref{fig_small_mean1} between the cumulants obtained using (\ref{pevolshort}) and the large-$\ell$ approximation (\ref{small_mean2}) is explained in DISCUSSION. 

{The main panel of Fig.~\ref{fig_small_mean1} shows that the mean is very close to the stationary state at time $t \approx 5000$. However, as shown in the inset, the mean continues to change albeit very slowly until $t \approx 2 \times 10^5$ when the true stationary state is reached. A similar pattern is seen for variance (and allele frequency shown below in Fig.~\ref{fig_small_freq}).  These observations suggest that the dynamics of the mean can be divided into a short-term phase in which the mean approaches a value close to the new optimum and a long-term phase in which it reaches the exact stationary state. To explore fast evolution, we will concentrate in the following on the short-term phase.}

\noindent{\bf Dynamics of the allele frequencies:} From (\ref{pevolshort}) \tc{black}{for allele frequency dynamics}, we first note that for $\Delta c_1(0) < 0$, {\it all}  the $+$ allele frequencies increase in the short-term phase (note that so far, we have argued that (\ref{pevolshort}) holds for small-effect loci only but in the next section, we will find this to be true for large-effect loci also). This simple result has potentially useful application as explained in DISCUSSION. 

Using (\ref{small_mean}) in (\ref{pevolshort}), the  time dependence of the allele frequency for small deviations in the phenotypic optimum can be obtained analytically \citep{Chevin:2008,Jain:2015}:
\be
p_i(t)=\frac{p_i(0)}{p_i(0)+ q_i(0) \exp\left(-\frac{z_f \gamma_i (1-e^{-s \ell {\bar \gamma}^2 t})}{\ell {\bar \gamma}^2}\right)}~.
\label{freqsmall}
\ee 
Thus in the directional selection model, the allele frequency reaches the stationary state over a time period $\sim (s \ell {\bar \gamma}^2)^{-1}$. These shifts in allele frequency are small to moderate, as verbally predicted by several authors (e.g. \citet{Pritchard:2010}). However, this directional behavior may change in the long-term period, as we will discuss next.

The allele frequency dynamics in the full model and directional selection model are compared for some loci in Fig.~\ref{fig_small_freq}, and we see that while the two match well at short times, in the full model, the allele frequency at large times can vary in a {\it nonmonotonic} fashion (see inset of Fig.~\ref{fig_small_freq}) and approaches a stationary state value given by the solution of (\ref{sseqn}).  The long-time dynamics of the allele frequency can be understood by solving (\ref{pevoleqnf}) with zero mean deviation and the initial condition given by the stationary state solution of the directional selection model \citep{Jain:2015}. A succinct way of expressing such a prescription is to write 
\be
\Delta c_1(t)=\Delta c_1^*+\Delta c_1(0) e^{-s c_2(0) t}
\label{meanmodi}
\ee
in (\ref{pevoleqnf}). Numerical integration of the resulting equation matches well with the full model as shown in Fig.~\ref{fig_small_freq}. An analytical understanding of the long-time dynamics can also be obtained as described in  Appendix~\ref{app_freq}.

\bigskip
\centerline{DYNAMICS WHEN MOST EFFECTS ARE LARGE}
\bigskip

We now turn to the situation when most effects are large (${\bar \gamma} \gg {\hat \gamma}$).  
Due to (\ref{icss}), the initial allele frequencies are close to either zero or one. When the final mean deviation is close to zero, either the final allele frequency stays close to the initial one or sweeps to fixation as shown in Fig.~\ref{fig_large_freq1}. 
Thus, unlike in the case where most effects are small, \tc{black}{in the full model} the factor $q_i-p_i$ in the last two terms on the RHS of  (\ref{pevoleqnf}) is not negligible. However, the directional selection model defined by (\ref{pevolshort}) describes the short-time dynamics when most effects are large as explained below. 

Since we are dealing with the case where ${\hat \gamma} \ll \gamma _i$, we can ignore the effect of mutations. Furthermore, as $|q_i-p_i|$ can not exceed one, (\ref{pevoleqnf}) shows that as long as $2 |c_1(t)-z_f| \equiv 2 |\Delta c_1(t)| \gg \gamma_i$, the second term on the RHS can be neglected to yield (\ref{pevolshort}) for the allele frequency at the $i$th locus. Therefore we expect the directional selection model to apply at short times. But as time progresses, the mean deviation decreases in magnitude and the above condition for the validity of the directional selection model fails. 

Using the initial frequency given by (\ref{icss}) for $\gamma_i \gg {\hat \gamma}$, equation (\ref{app_c1p1}) for the mean  yields 
\bea
c_1(t) = \sum_{i=1}^\ell \gamma_i - \sum_{i \in {P}^+}\frac{2 \gamma_i}{1+ \frac{p_i^{+}(0)}{q_i^+(0)}  e^{{\beta \gamma_i}/{\bar \gamma}}}-\sum_{i \in P^-} \frac{2 \gamma_i}{1+ \frac{p_i^{-}(0)}{q_i^-(0)}  e^{\beta {\gamma_i}/{\bar \gamma}}} ~,
\eea
where $P^\pm$ refers to the set of loci with initial frequencies $p_i^\pm (0)$ given in (\ref{icss}). From symmetry considerations, we expect the number of loci with initial frequency above and below one half to be equal. Furthermore, for $\gamma_i \gg {\hat \gamma}$, the initial allele frequency given by (\ref{icss}) can be approximated as $p_i^-(0) \approx (\frac{{\hat \gamma}}{2 \gamma_i})^2, p_i^+(0)=1-p_i^-(0) $. We thus arrive at
\be
c_1(t) \approx  \sum_{i=1}^\ell \gamma_i - \sum_{i \in {P}^+}\frac{2 \gamma_i}{1+ \left(\frac{2 \gamma_i}{\hat \gamma} \right)^2 e^{{\beta \gamma_i}/{\bar \gamma}}}-\sum_{i \in P^-} \frac{2 \gamma_i}{1+ \left(\frac{\hat \gamma}{2 \gamma_i} \right)^2 e^{\beta {\gamma_i}/{\bar \gamma}}}~.
\label{c1lar1}
\ee
Then, as in the last section, approximating the sums in (\ref{c1lar1})  by integrals for large $\ell$, we get
\be
{c_1(t)}={\ell {\bar \gamma}} (1 - I_l^+ -I_l^-) ~,
\label{c1lar2}
\ee
where
\bea
I_l^+ &=& \tc{black}{\frac{1}{2 \bar \gamma} \int_{\hat \gamma}^\infty d\gamma \frac{2 \gamma e^{-\gamma/{\bar \gamma}}}{1+ \left(\frac{2 \gamma}{\hat \gamma} \right)^2 e^{{\beta \gamma}/{\bar \gamma}}}} \\
&=& \int_{2/\alpha}^\infty dx ~\frac{x e^{-x}}{1+ (\alpha x)^2 e^{\beta x}} \label{Il+} ~,
\eea
and
\bea
I_l^- &=& \tc{black}{\frac{1}{2 \bar \gamma} \int_{\hat \gamma}^\infty d\gamma \frac{2 \gamma e^{-\gamma/{\bar \gamma}}}{1+ \left(\frac{\hat \gamma}{2 \gamma} \right)^2 e^{{\beta \gamma}/{\bar \gamma}}}} \\
&=& \int_{2/\alpha}^\infty dx ~\frac{x e^{-x}}{1+ (\alpha x)^{-2} e^{\beta x}} \label{Il-} ~.
\eea
In the above equations, the parameter $\alpha=2 {\bar \gamma}/{\hat \gamma}$. 

\noindent{\bf {Qualitative patterns:}} In the directional selection model, as (\ref{betaeqn}) and (\ref{cumshort}) hold irrespective of whether most effects are small or large, the mean deviation $\Delta c_1(t)$ and the 
allele-frequency-dependent variable $\beta$ increase with time as discussed in the last section, and the variance is given by $c_2=-\ell {\bar \gamma}^2 (I_l^{+'} +I_l^{-'})$. However, unlike when most effects are small, here the rate of change of variance, ${\dot c}_2= s {\ell} {\bar \gamma}^3 \Delta c_1 (I_l^{+''} +I_l^{-''})$ is not a monotonic function of time; this is because the sum $I_l^{+''} +I_l^{-''}$ is negative when $\beta$ is small and positive for larger $\beta$. Thus the directional selection model predicts that initially the variance {\it increases} with time and then decreases towards the stationary state at the new phenotypic optimum.


\noindent{\bf {Quantitative dynamics:}} 
Equation (\ref{c1lar2}) along with (\ref{betaeqn}) forms a closed system of equations and allows us to find the mean. The integrals $I_l^{\pm}$ are estimated in Appendix~\ref{app_large}; we find that the integral $I_l^+$ does not contribute substantially to the mean {since it includes the loci with initial frequency close to one and we are dealing with the case when the initial mean deviation is negative (see also (\ref{DSlarge}) below),} and $I_l^-$ is well approximated by $J_l$ given by (\ref{Jldefn}).  When the initial deviation from the phenotypic optimum is small, using (\ref{app_lar_sdevM2}) and (\ref{app_Jlprime}), we find that at large times, the mean deviation decreases exponentially fast, 
\bea
\Delta c_1(t) &\sim& e^{- s \ell {\bar \gamma}^2 (\ln \alpha)^{-1} \rho (\ln \rho)^2 t} ~,\\
&=&e^{- s z_f {\bar \gamma} (\ln \alpha)^{-1}  (\ln \rho)^2 t}~,
\label{large_mean2}
\eea
where $\rho=z_f/(\ell {\bar \gamma})$ {is the phenotypic shift relative to the maximum mean magnitude}. The above result shows that time scale over which the mean approaches the stationary state depends weakly on the number of loci unlike the case when most effects are small (see (\ref{small_mean})). However as the mean of the distribution is large here, the relaxation in the directional selection model occurs faster than in the small-effects case.  

When most effects are large, the variance changes considerably even when the relative phenotypic shift is quite small as Fig.~\ref{fig_small_meanL1} shows. 
The peak variance $c_2^{(\textrm{max})}$ can be estimated by the stationary state variance in the directional selection model. At large times, using (\ref{app_lar_svar2}) and (\ref{app_Jlprime}), we find that 
\be
c_2^{(\textrm{max})} = \ell {\bar \gamma}^2 (\ln \alpha)^{-1} \rho (\ln \rho)^2
\label{c2max}
\ee
and therefore $c_2^{(\textrm{max})}/c_2(0) \sim \alpha^2 \gg 1$. \tc{black}{This result can also be seen by noting that the allele frequencies at loci with initial frequency close to zero shift to intermediate values at short times. Then from (\ref{vardefn}), the maximum variance is obtained when the allele frequency is close to one half leading to $c_2^{(\textrm{max})} \sim \ell {\bar \gamma}^2$.} 

The dynamics of the allele frequency at long times are discussed in Appendix~\ref{app_freq} and at short times in the next section.

\bigskip
\centerline{WHEN DO SELECTIVE SWEEPS OCCUR?}
\bigskip

Below we obtain a criterion on the size of effects for which the allele frequency \tc{black}{at a major locus} can sweep. 
When $p_i(0) \approx 0$, at short enough times, keeping only the lowest order terms in $p_i$ on the RHS of the full model (\ref{pevoleqnf}) and neglecting the effect of mutations, we get
\be
{\dot p}_i \approx - \frac{s \gamma_i p_i}{2}  (2 \Delta c_1(0)+\gamma_i) ~.
\label{DSlarge}
\ee
When the initial mean deviation $\Delta c_1(0) < 0$, the above equation shows that only the loci with effect $ \gamma_i < 2 |\Delta c_1(0)|$ can potentially sweep since their frequency increases with time. On repeating the above analysis for loci with initial frequency close to one, we find that the frequency at such loci does not sweep to zero. 
Thus, when $\Delta c_1(0) < 0$, only the loci with small initial frequency and effect $< 2 |\Delta c_1(0)|$ are likely to undergo large changes in the frequency. 
These criteria are necessary but as Fig.~\ref{fig_large_freq1} shows, they are not sufficient for a selective sweep to occur.

For the major loci that meet the above necessary conditions for selective sweeps, the second and the third term on the RHS of \tc{black}{the full model} (\ref{pevoleqnf}) can be neglected in comparison to the first term thereby reducing (\ref{pevoleqnf}) to the directional selection model (\ref{pevolshort}). For the rest of the (major) loci, the allele frequency changes are not appreciable and the frequencies remain close to zero or one, a solution satisfied by (\ref{pevolshort}). \tc{black}{Let ${\tilde t}$ denote the time when the allele frequencies evolving according to the directional selection model equilibrate. Figure~\ref{fig_large_freq1} shows that the  mean deviation obtained using the full model is close to zero at time ${\tilde t}$.} 
Then for subsequent times $t > {\tilde t}$, we can ignore mutations and set $\Delta c_1=0$ in the full model (\ref{pevoleqnf}) to obtain \citep{Jain:2015}
\be
{\dot p}_i= \frac{s \gamma_i^2}{2} p_i q_i(2 p_i-1) ~,~ t > {\tilde t}~.
\label{tlarDS}
\ee
{The above differential equation is subject to the initial condition  $p_i({\tilde t})$ which is the allele frequency in the steady state of the directional selection model. }

From the definition (\ref{app_omegadefn}) of $\beta$, we have
\tc{black}
{\be
\frac{p_i({\tilde t})}{1-p_i({\tilde t})}=\frac{p_i(0)}{q_i(0)} \exp\left(\frac{\gamma_i {\tilde \beta}}{\bar \gamma} \right)
\ee}
which simplifies to give the frequency $p_i({\tilde t})$ as
\bea
p_i({\tilde t}) =\left[1+ \frac{q_i(0)}{p_i(0)} \exp\left(-\frac{\gamma_i {\tilde \beta}}{\bar \gamma} \right)\right]^{-1}~,
\eea
where ${\tilde \beta}\equiv \beta(\tilde t)$. It is clear from (\ref{tlarDS}) that if the frequency $p_i({\tilde t})$ is greater than one half, the allele frequency at the $i$th locus will increase monotonically towards one. Thus we predict that the allele frequency at a locus would sweep when $p_i({\tilde t}) > 1/2$. {Combining this criterion with the necessary condition mentioned above, we find that a sweep can occur when the effects lie in the following range:}
\be
\frac{\bar \gamma}{\tilde \beta} \ln \left(\frac{q^-_i(0)}{p^-_i(0)} \right) < \gamma_i < 2 z_f. 
\label{critss}
\ee
This expectation is indeed seen to be consistent with the numerical solution of the full model as shown in Fig.~\ref{fig_large_freq1} except for one case. 

\noindent{\bf When most effects are small:} While considering the dynamics of mean and variance when most effects are small in an earlier section, we ignored the contribution of (few) loci that have large effect. However the results obtained earlier can be utilized to understand the dynamics of the allele frequency at major loci also. As (\ref{betaS}) shows, the auxiliary parameter in the stationary state of the directional selection model is ${\tilde \beta}=z_f/(\ell {\bar \gamma})$. Using this in (\ref{critss}), we find that a selective sweep can occur at a major locus if its effect 
\bea
2 z_f > \gamma_i &>& \frac{\ell \bar \gamma^2}{z_f} \ln \left(\frac{q^-_i(0)}{p^-_i(0)} \right)   ~,~\\
&=& \frac{c_2(0)}{|\Delta c_1(0)|}  \ln \left(\frac{q^-_i(0)}{p^-_i(0)} \right) ~,
\eea
which matches the criterion (26b) of \citet{Chevin:2008}. Furthermore, using (\ref{icss}), we find that 
\be
\frac{2 c_2(0)}{|\Delta c_1(0)|}  \ln \left(\frac{2 \gamma_i}{{\hat \gamma}} \right) < \gamma_i < 2 z_f~.
\label{crit_sma}
\ee
For the parameters in Fig.~\ref{fig_small_mean1} where the average effect size ${\bar \gamma}=0.04$ and the threshold effect ${\hat \gamma}=0.128$, the above criterion requires that the effect size at a major locus be $> 0.81$. For exponentially distributed effects, as is assumed in this article, the probability for the effect size to exceed this criterion is extremely low ($\sim 10^{-9}$ for the parameters in Fig.~\ref{fig_small_mean1}) and therefore selective sweeps at major loci are prevented when most effects are small \citep{Chevin:2008,Pavlidis:2012}. 

\noindent{\bf When most effects are large:} The criterion for short-time sweeps is more involved in this case; using  (\ref{icss}) in (\ref{critss}), we obtain 
\bea
2 z_f > \gamma_i &>&  \frac{2 \bar \gamma}{\tilde \beta} \ln \left(\frac{2 \gamma_i}{{\hat \gamma}} \right)~,~ \label{largesweep} \\
&\approx&  \frac{2 \bar \gamma}{ \ln \left(\frac{ 2 {\bar \gamma}}{\hat \gamma} \right)} \ln \left(\frac{\ell {\bar \gamma}}{z_f} \right) \ln \left(\frac{2 \gamma_i}{{\hat \gamma}} \right)~,\label{crit_lar}
\eea
where we have used (\ref{app_betastar}) for $\tilde \beta$ in the last expression. 
For the parameters in Fig.~\ref{fig_small_meanL1}, using ${\tilde \beta \approx 2.32}$ in (\ref{largesweep}), we find that for a selective sweep to occur, $\gamma_i > 1.04$ is required whose probability is $\approx 0.024$. In other words, we 
expect as many as {$0.024 \times 200 \approx 5$} sweeps to occur for these parameter values over the time scale the mean deviation approaches zero.

\bigskip
\centerline{DISCUSSION}
\bigskip

\noindent{\bf Dynamical regimes in the full model:}  Figure~\ref{fig_large_freq1} shows an example of the dynamics of the mean and allele frequencies when an infinitely large population in linkage equilibrium and equilibrated to a phenotypic optimum is suddenly shifted to a new optimum. We observe a {\it short-term phase} $(t \lesssim 10^3)$ where the mean deviation increases quickly to a value close to zero followed by a {\it long-term phase} $(10^3 < t < 10^5)$ where  the mean deviation undergoes minor changes at a slow pace. 
The former phase where the adaptation process is rapid is the focus of this article since: (i) much of the action happens in this early phase (for example, in Fig.~\ref{fig_large_freq1}, the allele frequencies at $13$ out of $20$ loci were close to the stationary state at the end of the short-term phase), (ii) these time scales are experimentally observable, and very importantly, (iii) all $+$ alleles increase in frequency in a synergistic manner.

\noindent{\bf Directional selection model:} Although the full model defined by (\ref{pevoleqnf}) can be investigated numerically \citep{Vladar:2014}, as the allele frequency dynamics at a single locus are related to all the other $\ell-1$ loci via the phenotypic mean (first term on the RHS of (\ref{pevoleqnf})), it is difficult to obtain analytical results. Moreover, the full model results in equations for the cumulants that do not close, and therefore one resorts to an often unjustified and uncontrolled approximation of setting all the cumulants above a certain number to zero {(see \citet{Barton:2009} for a discussion of this approach)}. 
This approximation, however, works quite well when most effects are small \citep{Jain:2015} but fails completely when most effects are large 
(see supplemental file~\ref{app_cumhie} for a detailed explanation). 

Here we have devised an approximate model (directional selection model) defined by (\ref{pevolshort}) in which the allele frequencies are coupled but nonetheless, one can make analytical progress without truncating the cumulants arbitrarily. The central equations in the directional selection model are (\ref{app_c1p1}) and (\ref{betaeqn}) that relate the mean deviation and an auxiliary parameter $\beta$ which is a function of a single allele frequency. As explained in Appendices~\ref{app_small} and \ref{app_large}, it is possible to obtain explicit expressions for the time dependence of $\beta$ and thereby that of the mean. Higher cumulants such as the variance can be then found using (\ref{cumshort}). 

\noindent{\bf {Distribution of effects:}} All the numerical examples shown in the figures use a single realization of the effect at a locus, {\it i.e.}, one set of effects $\{\gamma_i \}~,~i=1,...,{\ell},$ are chosen from an exponential distribution. For additive quantities such as mean and variance that are obtained by summing the contribution over a  number of loci, we expect that a single realization of effects is a good representative when the number of loci are sufficiently large (in physics literature, this concept is known as self-averaging \citep{Castellani:2005}). A similar idea has been used recently in a model of stabilizing selection with the same effects at all loci in which the phenotypic mean for one set of initial allele frequencies is approximated by the corresponding average \citep{Charlesworth:2013b}. 
 Figure~\ref{fig_infL} shows that the quantitative agreement between the results for the mean deviation obtained for a single realization of phenotypic effects for $\ell=50, 200, 400$ and the analytical result for large $\ell$ gets better as the number of loci increase.
 
\noindent{{\bf Main results:} The key conclusions of this article are discussed below and  summarised in Table~\ref{tab_summ}.}

\begin{table}[t]
\begin{center}
\begin{tabular}{|c|c|c|c|}
\hline
 & Mean  dynamics  & Variance  dynamics & Selective sweeps  \\
$$ &  determined by & characterised by & at major loci  \\
\hline
\hline
Small effects & Initial variance (\ref{small_mean}) & Small variations (\ref{app_varsma}) & Unlikely (\ref{crit_sma}) \\
Large effects & Effect size (\ref{large_mean2}) & Large variations (\ref{c2max}) & Probable (\ref{crit_lar}) \\
\hline
\end{tabular}
\caption{{Summary of the results for the dynamics of phenotypic mean, variance and allele frequency when the phenotypic optimum is suddenly shifted.}}
\label{tab_summ}
\end{center}
\end{table}

\noindent{\it Response of the mean to a sudden shift of the phenotypic optimum:} How does the time scale over which the mean reaches the stationary state depend on various factors such as the number of loci, initial phenotypic mean deviation and size of the effects? We find that the mean deviation becomes negligible on a time scale that is inversely proportional to the number of loci ($\ell$) when most effects are small but depends weakly on $\ell$ when most effects are large (cf. (\ref{small_mean}) and (\ref{large_mean2})). 
This difference in the behavior may be understood as follows: when most effects are small, there is sufficient genetic variation initially for selection to act on because a large number of loci are initially polymorphic. Therefore, with increasing $\ell$, \tc{black}{the equilibration time} decreases. In contrast, when most effects are large, the initial genetic variation is negligible and the adaptation dynamics are determined by the size of the effects. 
As the availability of large-effect loci depends on the {mean} of the effect distribution, it follows that the larger the mean, the faster the adaptation dynamics. 

Figure~\ref{fig_zeffS} shows that the relationship between the equilibration time and the initial mean deviation depends on the size of effects \tc{black}{(see also \citet{Gomulkiewicz:2010})}. When most effects are large, selective sweeps occur at some loci that result in drastic changes in the allele frequencies thus accelerating the approach to the phenotypic optimum. But when most effects are small, 
\tc{black}{the equilibration time} remains roughly constant  as selective sweeps do not occur but small changes in the allele frequencies at a large number of loci drive the adaptation process. 
Figure~\ref{fig_zeffS} also shows that when the number of loci contributing to a trait are the same, adaptation is faster when the effect size is larger, which makes intuitive sense. 

However one may also compare situations in which a large number of small effects control a phenotypic trait with the one in which few loci with large effects contribute to a trait. An example shown in Fig.~\ref{fig_Fewl} suggests that the adaptation at short times is faster when effects are small but at longer times, larger effects lead to rapid adaptation. This conclusion is borne out by our analysis as well: on comparing (\ref{small_mean}) and (\ref{large_mean2}), we find that for identical $z_f$ and $\ell {\bar \gamma}$, the time scales are inversely proportional to $\bar \gamma$ and therefore it takes shorter time to equilibrate in the large-effect case. 

An important omission in the above discussion is random genetic drift as we have considered infinitely large populations. Some progress in this regard has been made recently {\citep{Matuszewski:2015,Budova:2016}}; however, a detailed analysis is currently not available and would be interesting to consider in the future. 

\noindent{\it Dynamics of the variance and allele frequencies:} Besides the dynamics of the mean, we also studied the time dependence of the variance. We find that at short times, the variance does not vary much and decreases monotonically with time when most effects are small. \tc{black}{This behavior can be explained by the fact that the allele frequencies show small to moderate changes.} In the opposite case, the variance changes considerably and its variation is nonmonotonic in time - it increases at short times and then decreases to the stationary state value. \tc{black}{The selective sweeps at major loci are responsible for this effect.} The rate of change of average fitness given by (\ref{app_fit}) receives contributions from the square of the mean deviation and the rate of change of variance. Thus when most effects are large, at short times, the average fitness increases more slowly in the short-term phase than at longer times \citep{Gomulkiewicz:2010}.

The directional selection model describes the bulk of the mean dynamics in the short-term phase. 
Therefore at late times where the mean deviation is close to zero, the allele frequency at each locus evolves independently (the first term on the RHS of (\ref{pevoleqnf}) is zero) and the dynamics are much slower as described in Appendix~\ref{app_freq}. At short times, the allele frequency dynamics involve the same time scales as for the mean and variance.

\noindent{\bf Applications:} The analysis of our model relates to various important questions that are currently discussed in evolutionary genetics.                 

\noindent{\it Rapid adaptation:} The approximations we have presented here describe the short-term evolution of a phenotypic trait after a sudden environmental change very well. We have shown that the mean of a phenotypic trait may respond very fast after an environmental shift. In the case that most effects are small, this is possible because the time to the new optimum is inversely proportional to the number of loci controlling the trait. In the opposite case of mostly large effects, rapid fixations (leading to selective sweeps) may produce fast phenotypic responses. In the examples of fast adaptation mentioned in INTRODUCTION both of these extremes and combinations thereof have occurred.

\noindent{\it QTL and selective sweep mapping:} Selective sweep mapping has been used to dissect QTL 
\citep{Svetec:2011,Rubin:2012,Axelsson:2013,Wilches:2014}. However, the success of this approach was varied. Nonetheless, there seems to be a tendency that it worked better in domesticated than in natural populations 
\citep{Rubin:2012,Axelsson:2013}\tc{black}{, probably due to the action of artificial selection during domestication. In artificial selection, the shift to the new optimum $z_f$  may be larger than under natural conditions; our criterion (\ref{critss}) thus indicates an enhancement of sweeps in domesticated populations.}

More generally, our analysis provides also some new insights into the question whether selective fixations (and thus sweeps) occur at QTL. While \citet{Chevin:2008} predicted that the probability of selective sweeps is extremely low at QTL (based on a model with one major locus and infinitely many minor loci), others have found sweeps at appreciable frequencies using simulations of various multi-locus models \citep{Pavlidis:2012,Wollstein:2014}. The prediction of Chevin and Hospital is consistent with our study for mostly small-effect loci and small shifts in the phenotypic optimum. 

\noindent{\it Testing for fast polygenic adaptation:} Using standardized frequencies, one can construct tests to detect SNPs that deviate strongly from neutral population structure (e.g. \citet{Coop:2010}). However, this approach is only working if there exist relatively large extended gradients of ecological variables, which may not be the case in rapid adaptation. Shortly after a population occupies a new habitat, we expect that the allele frequency shifts between the parental and derived populations are relatively small. This also means that available software, such as BayeScan-like methods \citep{Foll:2008,Riebler:2008}, is not able to detect significant frequency shifts for individual SNPs between populations. Therefore, for detecting small allele frequency shifts after environmental changes in fast adapting populations, it may be best to consider the frequency shifts of alleles simultaneously at all loci involved (instead of individual SNPs). Our results suggest that this may be a promising approach, as all + alleles shift their frequencies in the same direction in the short-term phase, which should increase the power of the test. {In human population genetics, this approach has been used to analyze recent polygenic adaptation for instance of height \citep{Turchin:2012}.}

\noindent{\bf Acknowledgements:} We thank the ICTS, Bangalore, for their hospitality during the Second Bangalore School on Population Genetics and Evolution (ICTS/Prog-popgen/2016/01) that facilitated our discussions. Furthermore, we are grateful to Graham Coop and two anonymous reviewers who made valuable suggestions to improve the paper. The research of WS was supported by the German Research Foundation DFG (grant Ste 325/17-1 from the Priority Program 1819).

\clearpage

\begin{center}
{\bf {\large APPENDIX}}
\end{center}

\appendix

\renewcommand{\thesection}{\Alph{section}}
\numberwithin{equation}{section}

\section{Cumulants of the effects and their dynamics}
\label{app_cum}

{The effect $e_i$ at the locus $i$ is distributed according to a Bernoulli distribution 
\be
p(e_i)=p_i \delta_{e_i,\frac{\gamma_i}{2}} + q_i \delta_{e_i,-\frac{\gamma_i}{2}} ~.
\ee
The generating function of the distribution of the effects at the $i$th locus is given by 
\be
F_i(\xi)=\sum_{e_i} p(e_i) e^{-e_i \xi}=p_i e^{-\frac{\gamma_i}{2} \xi}+q_i e^{\frac{\gamma_i}{2} \xi}~.
\ee
The logarithm of the generating function defines the $n$th cumulant $c_n^{(i)}$ at locus $i$ as  \citep{Sornette:2000}
\be
\ln F_i(\xi)
=\sum_{n=1}^\infty c_n^{(i)} \frac{(-\xi)^n}{n!} = \frac{\xi \gamma_i}{2}+\ln \left(q_i +p_i e^{-\gamma_i \xi}  \right) ~.
\label{cumgenfn}
\ee
The $n$th cumulant is then obtained as 
\be
c_n^{(i)}= (-1)^n \frac{d^n \ln F_i(\xi)}{d \xi^n} \Big|_{\xi=0} ~.
\ee 
In linkage equilibrium, since the cumulants for the $n$-locus problem are additive, we get
\be
c_n= 2 \sum_{i=1}^\ell c_n^{(i)} ~,
\label{sumcum}
\ee
where the factor of $2$ on the RHS takes care of diploidy. 
Using the last three equations above, we find that the mean and variance are given by (\ref{meandefn}) and (\ref{vardefn}), respectively. The skewness $c_3$ which is a measure of asymmetry of a distribution can also be found and given by
\be
c_3 = 2 \sum_{i=1}^\ell \gamma_i^3 (q_i-p_i) p_i q_i ~.\label{app_skewdefn}
\ee

{
\section{Quasi-linkage equilibrium approximation}
\label{app_ld}}


For the additive genotype-phenotype map, the trait $z=\sum_{i=1}^\ell \gamma_i (X_i+ X^*_i-1)$ where $X_i,X^*_i=1 (0)$ if the $+ (-)$ allele is present at the locus $i$. 
Following \citet{Barton:1991}, we first find the association coefficients by writing $X_i= \langle X_i \rangle + \delta_i, X^*_i= \langle X^*_i \rangle + \delta^*_i$ where $\langle X_i \rangle =\langle X^*_i \rangle=p_i$ and expanding the relative fitness to lowest orders in $s$, 
\bea
\frac{w(\{X,X^*\})}{\overline w} &=& \frac{1-(s/2) (z-z')^2}{{\overline w}} \\
& \approx & 1+ \sum_{i=1}^\ell a_{i,\phi} (\delta_i-\langle \delta_i \rangle) +  \sum_{i=1}^\ell a_{ii,\phi} (\delta_i^2-\langle \delta_i^2 \rangle)  \nonumber \\
&+& \sum_{i=1}^\ell a^*_{i,\phi} (\delta_i^*-\langle \delta_i^* \rangle) + \sum_{i=1}^\ell a^*_{ii,\phi} (\delta_i^{*2}-\langle \delta_i^{*2} \rangle)  \nonumber \\
&+& \sum_{i < j} a_{ij,\phi} (\delta_i \delta_j -\langle \delta_i \delta_j \rangle) +\sum_{i < j} a^*_{ij,\phi} (\delta_i^* \delta_j^* -\langle \delta_i^* \delta_j^* \rangle) \nonumber \\
&+&  \sum_{i,j=1}^\ell a_{i,j} (\delta_i \delta_j^* -\langle \delta_i \delta_j^* \rangle) ~.
\eea
In the above equation, the average is taken with respect to the joint distribution of the frequencies at all the loci and the coefficients 
\be
a_{i,\phi}=-s (c_1-z') \gamma_i,~a_{ii,\phi}=-(s/2) \gamma_i^2,~a_{ij,\phi}=a_{i,j}=-s \gamma_i \gamma_j ~.
\ee

The average allele frequency change after selection (and recombination) is given by (see (10a), \citet{Barton:1991})
\be
\dot p_i = a_{i,\phi} \langle \delta_i^2 \rangle + a_{ii,\phi} \langle \delta_i^3 \rangle + \sum_{j \neq i} a_{j,\phi} \langle \delta_i \delta_j \rangle + \sum_{j \neq i} a_{jj,\phi} \langle \delta_i \delta_j^2 \rangle +\sum_{j \neq k} a_{kj,\phi} \langle \delta_i \delta_j \delta_k \rangle ~.
\label{app_avgp}
\ee
All but the first two terms on the RHS vanish in linkage equilibrium (LE); here we focus on the lowest order corrections to LE which are contained in the third and fourth terms. Using $X_i^2=X_i$ and $\langle \delta_i \rangle=0$, it is easy to show that $\langle \delta_i^2 \rangle=p_i q_i, \langle \delta_i \delta_j^2 \rangle=(1-2 p_j) \langle \delta_i \delta_j \rangle$. Inserting these expressions in (\ref{app_avgp}), we verify that the first two terms on the RHS match the corresponding ones in (\ref{pevoleqnf}). The two-locus linkage disequilibrium evolves according to (see (12), \citet{Barton:1991})
\be
\dot {\langle \delta_i \delta_j \rangle}=-r_{ij} \langle \delta_i \delta_j \rangle +2 a_{ij,\phi} \langle \delta_i^2 \rangle \langle \delta_j^2 \rangle+ \textrm{higher order terms} ~,i \neq j ~,
\ee
where $r_{ij}$ denotes the recombination rate between the loci $i$ and $j$ and we have dropped terms that are nonlinear in the parameters $s$ and $r$ on the RHS of the above equation. Then at quasi-linkage equilibrium (QLE) where the LHS is of ${\cal O}(s^2)$ or higher, we have
\bea
\langle \delta_i \delta_j \rangle_{QLE} &\approx& \frac{2 a_{ij,\phi}}{r_{ij}} \langle \delta_i^2 \rangle_{\textrm{LE}}\langle \delta_j^2 \rangle_{\textrm{LE}} \\
&=& -\frac{2 s \gamma_i \gamma_j}{r_{ij}} (p_i q_i p_j q_j) |_{\textrm{LE}} ~,
\eea
where the allele frequencies in the last equation are in LE. 

Using the above result in (\ref{app_avgp}) and assuming that all the recombination rates are equal to $r$, we find that within QLE, the allele frequency obeys the following equation:
\be
\dot p_{i} \approx -s \gamma_i p_i q_i  (c_1-z') \left(1- \frac{s}{r} c_{2,\textrm{LE}} \right)-\frac{s \gamma_i^2}{2} p_i q_i(q_i-p_i) \left(1- \frac{s}{r} \frac{c_{3,\textrm{LE}}}{\gamma_i} \right)
\ee
where $c_{2,\textrm{LE}}$ and $c_{3,\textrm{LE}}$ are the variance and skewness, respectively, in linkage equilibrium. Thus, as expected, the corrections to the model defined by (\ref{pevoleqnf}) are of order $s/r$ which can be neglected when selection and linkage are weak. 


\section{Directional selection model}
\label{app_genl}

{The dynamics of the allele frequency $p_i$ are described by (\ref{pevolshort}) where the effective selection coefficient $S_i=-s \gamma_i \Delta c_1$ depends on all the allele frequencies through the mean $c_1$. This property makes it difficult to solve for the allele frequencies; however, we note that for two arbitrary loci $i$ and $m$, (\ref{pevolshort}) gives
\be
\frac{\dot p_i}{\dot p_m}=\frac{dp_i}{dp_m}= \frac{S_i p_i q_i}{S_m p_m q_m} ~,
\label{varrr}
\ee
which, on using the identity $d \left[\ln (p/q) \right]= dp/ (p q)$, simplifies to
\be
\frac{d \ln \varrho_i}{d \ln \varrho_m}= \frac{\gamma_i}{\gamma_m}~,~i \neq m ~,
\label{varr}
\ee
where $\varrho_i(t)= {p_i(t)}/{q_i(t)}$. As the above equation does not involve the phenotypic mean, it can be easily solved to give 
\be
\varrho_i(t)= \varrho_i(0) \left( \frac{\varrho_m(t)}{\varrho_m(0)}\right)^\frac{\gamma_i}{\gamma_m} ~.
\label{equiv}
\ee
We remark that the simple idea used in (\ref{varrr}) to eliminate the global variable $\Delta c_1$ is quite  similar to that employed in single-locus directional selection models to get rid of the average fitness (see, for example, p. 73, \citet{Charlesworth:2010}).}

Equation (\ref{equiv}) is useful since it allows us to express the mean $c_1(t)$ in terms of a single allele {frequency, say, at locus $m$}. Using (\ref{equiv}) in (\ref{meandefn}), we can write the mean as
\bea
c_1(t) &=&  \sum_{i=1}^\ell \gamma_i \left(\frac{\varrho_i-1}{\varrho_i+1} \right)\\
&=& \sum_{i=1}^\ell \gamma_i -\sum_{i=1}^\ell \frac{2 \gamma_i}{1+\varrho_i(0) \left(\frac{\varrho_m(t)}{\varrho_m(0)} \right)^{\frac{\gamma_i}{\gamma_m}}} ~,
\label{app2_c1p1}
\eea
which simplifies to give (\ref{app_c1p1}). In the above equation, 
\bea
{\beta (t)} = \frac{\bar \gamma}{\gamma_m} \ln \left(\frac{\varrho_m(t)}{\varrho_m(0)} \right)= \frac{\bar \gamma}{\gamma_m} \ln \left( \frac{p_m(t)}{p_m(0)}\frac{q_m(0)} {q_m(t)}  \right)~.
\label{app_omegadefn}
\eea
Note that in the expression (\ref{app2_c1p1}), although the dynamics of the mean deviation are determined by that of the allele frequency $p_m$, the rest of the $\ell-1$ frequencies are not redundant as the mean depends on their initial state also.

Furthermore, from the evolution equation (\ref{pevolshort}) for the allele frequency, we also have 
${\dot \beta} = -s {\bar \gamma} \Delta c_1$, where the mean is given by (\ref{app2_c1p1}); we have thus obtained a {\it closed} equation for $\beta(t)$. 
Using the result for $\beta(t)$ in (\ref{app2_c1p1}), the dynamics of the phenotypic mean can be calculated. This knowledge then allows us to find higher cumulants using (\ref{cumshort}) which is derived below.

To obtain the evolution equation (\ref{cumshort}) for the $n$th cumulant, we take the derivative with respect to time in (\ref{cumgenfn}) and use (\ref{pevolshort}) for the allele frequency dynamics to obtain \citep{Burger:1991}
\be
\sum_{n=1}^\infty {\dot c}_n^{(i)} \frac{(-\xi)^n}{n!} = \frac{s \gamma_i p_i q_i \Delta c_1 (1-e^{- \gamma_i \xi})}{q_i +p_i e^{-\gamma_i \xi}}~.
\ee
But, on taking a derivative of  (\ref{cumgenfn}) with respect to $\xi$, after some simple algebra, we also have 
\be
\sum_{n=1}^\infty c_{n+1}^{(i)} \frac{(-\xi)^n}{n!} = - \frac{\gamma_i p_i q_i (1-e^{- \gamma_i \xi})}{q_i +p_i e^{-\gamma_i \xi}}~.
\ee
On comparing the last two equations and using (\ref{sumcum}), we arrive at the promised result (\ref{cumshort}). The rate of change of the logarithm of the average fitness can be found using the above results and given by
\be
\frac{d \ln {\bar w}}{dt}= \frac{\dot {\bar w}}{{\bar w}} =-\frac{s}{2} {\dot c}_2+c_2 (s \Delta c_1)^2~.
\label{app_fit}
\ee

\section{Cumulant dynamics when most effects are small}
\label{app_small}

Here we study the differential equation (\ref{betaeqn}) obeyed by $\beta$, 
\be
\dot \beta=-s \ell {\bar \gamma}^2 \left[1-\rho -2 I_s(\beta) \right]~,
\label{S_betaeqn}
\ee
where $\rho=z_f/{\ell {\bar \gamma}}$ and the integral 
\bea
I_s &=& \int_0^{{\hat \gamma}/{\bar \gamma} } dx ~\frac{x e^{-x}}{1+e^{\beta x}} \\
&=& \int_0^{\infty} dx ~\frac{x e^{-x}}{1+e^{\beta x}}-\int_{{\hat \gamma}/{\bar \gamma} }^\infty dx ~\frac{x e^{-x}}{1+e^{\beta x}}~.
\label{app_Is}
\eea
The first integral on the RHS is exactly solvable in terms of special functions and we have 
\bea
I_s^{(1)} &=& \int_0^{\infty} dx \frac{x e^{-x}}{1+e^{\beta x}} ~,
\label{Is1} \\
&=& {\Omega^2} \left[ \psi^{(1)} \left( \Omega+1/2\right) -  \psi^{(1)} \left( \Omega+1\right)  \right] ~, \\
&=& 1+ \frac{2 \psi^{(1)} (2 \Omega)- \psi^{(1)} (\Omega) }{2 \beta^2}~,
\eea
where $\Omega=(2 \beta)^{-1}$ and $\psi^{(1)} (z)$ is the second derivative of the logarithm of the gamma function ((6.4.1), \cite{Abramowitz:1964}). Since ${\bar \gamma} \ll {\hat \gamma}$ when most effects are small, the second integral on the RHS of (\ref{app_Is}) can be {estimated} by carrying out an integration by parts and we have
\be
I_s^{(2)} \approx \frac{{\tilde x} e^{-\tilde x}}{1+e^{\beta \tilde x}}~,
\ee
where $\tilde x={\hat \gamma}/{\bar \gamma}$. The above integral is thus exponentially small in $\tilde x$ and can be neglected in comparison to  $I_s^{(1)}$. 
Using the series expansion for the polygamma function $\psi^{(1)}(z)$ for small arguments ((6.4.10) and (6.4.11), \cite{Abramowitz:1964}), we obtain
\begin{numcases}
{I_s \approx}
\frac{1-\beta+\beta^3}{2}+...~ &,~~$\beta \ll 1$   \label{betass}\\
\frac{\pi^2}{12 \beta^2}+ {\cal O}(\beta^{-3})~&,~~$\beta \gg 1$~. \label{betasl}
\end{numcases}


As both the initial and stationary state frequencies are close to one half, the function $\beta(t)$ given by (\ref{app_omegadefn}) stays close to zero. Using the result (\ref{betass}) in the equation (\ref{S_betaeqn}) for $\beta(t)$, we get
\be
\frac{d \beta}{d \tau} \approx - \beta+ \beta^3 + \rho~,
\label{app_betaSS}
\ee
where $\tau=s \ell {\bar \gamma}^2 t$. As $\beta \ll 1$, ignoring the term $\beta^3$ on the RHS of the above equation and integrating, we obtain the zeroth order solution $\beta_0$  given by 
\be
 \beta_0(\tau)= \rho (1- e^{-\tau}) ~,
 \label{betaS}
 \ee
where we have used that $\beta(0)=0$. An approximate solution of (\ref{app_betaSS}) can be found iteratively by writing $\beta=\beta_0+\beta_1$ in (\ref{app_betaSS}) and retaining the leading order term in $\beta_1$ to obtain
\be
\frac{d \beta_1}{d \tau}+ \beta_1 \approx \beta_0^3 ~,
\ee
which yields $\beta \approx \beta_0 + \rho^{-1} \beta_0^4$. Using this result in (\ref{betass}), we get $2 I_s \approx 1- \beta_0+\beta_0^3 e^{-\tau}$. Equations (\ref{small_mean2}) and (\ref{cumshort}) then yield the first two cumulants to be 
\bea
\Delta c_1(\tau) &=& -z_f e^{-\tau} (1+ \rho^{-1} \beta_0^3) \label{app_mvsma1}\\
c_2(\tau) &=& \ell {\bar \gamma}^2 (1-3 \beta_0^2) ~. \label{app_mvsma2}
\eea


\section{Long-time dynamics of the allele frequency}
\label{app_freq}

When most effects are small, as shown in Fig.~\ref{fig_small_mean1}, the mean deviation is close to zero when $t \approx 10^4$. But the exact stationary state for the mean and the allele frequency (shown in Fig.~\ref{fig_small_freq}) is reached much later at $t \approx 10^5$. A similar qualitative behavior of the allele frequencies is seen when most effects are large in Fig.~\ref{fig_large_freq1}. To understand the slow dynamics at  long times when the effect size can be small or large, here, we study (\ref{pevoleqnf}) for the allele frequency dynamics in the full model approximating the mean deviation by its stationary state value $\Delta c_1^*$. We thus have 
\be
{\dot p}_i \approx -s \gamma_i p_i q_i  \Delta c_1^*-\frac{s \gamma_i^2}{2} p_i q_i(q_i-p_i) +\mu (q_i-p_i)~.
\label{app_pevoleqnf}  
\ee
An exact solution of the above equation requires solving a cubic equation which is possible to do but not particularly enlightening.

In order to find the behavior of the allele frequency close to the stationary state, we write $p_i(t)={ p}_i^*+y_i^{-1}(t)~,~y_i \gg 1$ in the above equation and obtain 
\be
-{\dot y}_i= -s \gamma_i^2 y_i^{-1}(t)+ { D}_1 - { D}_2 y_i(t)~,
\ee
where ${ p}_i^*$ is the exact solution of (\ref{sseqn}) with nonzero $\Delta c_1^*$ 
and the constants ${ D}_1$ and  ${ D}_2$ are functions of $\gamma_i, s, \mu, \Delta c_1^*$ . As we are interested in large times where the allele frequency is close to the stationary state, the first term on the RHS of the above equation can be neglected and we arrive at a first-order differential equation for $y_i$ which can be easily solved. For large times, we then obtain $p_i(t)-{ p}_i^* \sim e^{-{ D}_2 t}$, 
where 
\be
{ D}_2=s \gamma_i^2 \left[\frac{1}{2}+\left(\frac{{\hat \gamma}}{2 \gamma_i}\right)^2 -3 { p}_i^* (1-{ p}_i^*) + \frac{(1-2 { p}_i^*) \Delta c_1^*}{\gamma_i}\right]~.
\ee
Thus after the directional selection phase, the allele frequency at the locus with effect $\gamma_i$ reaches a stationary state over a time scale $\sim { D}_2^{-1}$. For small mean deviation from the optimum which results in negligible $\Delta c_1^*$, we can use (\ref{icss}) to obtain
\begin{numcases}
{ { D}_2 \propto} 
s {\hat \gamma}^2 &,~$\gamma_i \ll {\hat \gamma}$ \label{app_small_varLar1}\\
s \gamma_i^2 &,~$\gamma_i \gg {\hat \gamma}$ ~.
\label{app_small_varLar} 
\end{numcases}
The above equation shows that when most effects are small, the dynamics at long times are driven by mutations (recall that ${\hat \gamma}^2=8 \mu/s$) while in the opposite case, the effect size continues to play a role even at long times.


\section{Cumulant dynamics when most effects are large}
\label{app_large}

Consider first the integral $I_l^+$ defined by (\ref{Il+}) that includes the contribution from loci with initial frequency close to one and given by
\bea
I_l^+ &=& \int_{2/\alpha}^\infty dx ~\frac{x e^{-x}}{1+ (\alpha x)^2 e^{\beta x}} ~,\\
&\approx & \frac{1}{\alpha^{2}} \int_{2/\alpha}^\infty dx ~\frac{e^{-(1+\beta) x}}{x} ~,
\label{J11}
\eea
where $\alpha=2 {\bar \gamma}/{\hat \gamma}$ is large. 
The last integral can be written as  an exponential integral $E_1(z)= \int_1^\infty dt ~t^{-1} e^{-z t}$ and we have 
\bea
I_l^+ &=& \frac{1}{\alpha^{2}} E_1 \left[ \frac{2}{\alpha}(1+\beta)  \right] ~, \\
&\approx& -\frac{1}{\alpha^{2}} \ln \left[ \frac{2}{\alpha}(1+\beta) \right] ~,
\label{aJ_lpp_Ilmax}
\eea
on using that the argument of $E_1(z)$ is small \citep{Abramowitz:1964}. We next consider the integral  $I_l^-$ defined by (\ref{Il-}) that contains the contribution from loci with small initial frequencies, 
\bea
I_l^-&=& \int_{2/\alpha}^\infty dx ~\frac{x e^{-x}}{1+ (\alpha x)^{-2} e^{\beta x}} ~,\\
&=& \int_{2/\alpha}^\infty dx ~{x e^{-x}}+\int_{0}^{2/\alpha} dx ~\frac{x e^{-x}}{1+ (\alpha x)^{2} e^{-\beta x}} - J_l ~,\\
&\approx& 1- \left( \frac{2}{\alpha} \right)^2 - J_l~,
\label{app_Ilmin}
\eea
where we have used that $\alpha \gg 1$ to obtain the last expression and defined 
\be
J_l=\int_{0}^\infty dx ~\frac{x e^{-x}}{1+ (\alpha x)^{2} e^{-\beta x}}~.
\label{Jldefn}
\ee
As shown in Table~\ref{tab_Jl}, the integral $J_l$ decreases as $1/\alpha$ or slower. Thus, for large $\alpha$, equation (\ref{c1lar2}) for the mean gives 
\be
c_1(t) = \ell {\bar \gamma} (1-I_l^+-I_l^-) \approx \ell {\bar \gamma} J_l(\beta)~.
\label{app_c1l}
\ee
In the above equation, the integral $I_l^+$ does not contribute to the mean since, as discussed after (\ref{DSlarge}), the allele frequency at loci with high initial frequencies does not change significantly while the mean is evolving. 

The time dependence of $\beta$ is obtained from its evolution equation (\ref{betaeqn}), 
\be
\frac{d \beta}{d \tau} \approx  \rho-J_l(\beta) ~.
\label{app_betal}
\ee
The function $\beta$ increases with time and saturates to its steady state value ${\tilde \beta}$ where ${\dot \beta}=0$. Correspondingly, as Table~\ref{tab_Jl} indicates, the integral $J_l(\beta)$ is close to zero when $\beta \ll 1$ and approaches $\rho$ as $\beta \to {\tilde \beta}$. Thus at short times, we can neglect the term $J_l$ on the RHS of (\ref{app_betal}) and 
at large times, we can expand $J_l$ about $\rho$ to write (\ref{app_betal}) as ${\dot \beta} \approx -s \ell {\bar \gamma}^2 J_l'({\tilde \beta}) (\beta-{\tilde \beta})$. 
Denoting the time below and above which these two solutions are valid by $t_\times$ and setting $\beta(t_\times)=1$, we obtain
\begin{numcases}
{\beta \approx}
 s \ell {\bar \gamma}^2 \rho t&,~$t \ll t_\times$   \label{lar_sdev1}\\
{\tilde \beta}+{(1-{\tilde \beta}) e^{- \omega(t-t_\times)}}&,~$t \gg t_\times $  \label{lar_sdev2}
\end{numcases}
where $t_\times=(s \ell {\bar \gamma}^2 \rho)^{-1}, J_l({\tilde \beta})=\rho$ and $\omega=s \ell {\bar \gamma}^2 J_l'({\tilde \beta})$. Using this result in (\ref{app_c1l}), we find the mean deviation to be 
\begin{numcases}
{-\Delta c_1(t) \approx}
 z_f &,~$t \ll t_\times$   \label{lar_sdevM1}\\
 {\ell {\bar \gamma}} J_l'({\tilde \beta}) ({\tilde \beta}-1) e^{- s \ell {\bar \gamma}^2 J_l'({\tilde \beta}) (t-t_\times)}&,~$t \gg t_\times $ ~.\label{app_lar_sdevM2}
\end{numcases}
Using the above expressions in (\ref{cumshort}), we find that the variance is given by
\begin{numcases}
{c_2(t) \approx}
0 &,~$t \ll t_\times$   \label{lar_svar1}\\
 {\ell {\bar \gamma}^2} [J_l'({\tilde \beta})- ({\tilde \beta}-1) J_l''({\tilde \beta}) e^{- \omega (t-t_\times)}]&,~$t \gg t_\times $ ~.\label{app_lar_svar2}
\end{numcases}

In order to express the mean and the variance in terms of the model parameters, we first need to determine ${\tilde \beta}$ which is a solution of the equation $J_l({\tilde \beta})=\rho$. The integral $J_l(\beta)$ is not exactly solvable but we can estimate it by a saddle point method (see (\ref{saddle}) below) \citep{Arfken:1985}. Denoting the integrand of $J_l$ in (\ref{Jldefn}) by $j(x)$ ({\it i.e.}, $J_l=\int_0^\infty j(x) dx$) and the maximum of $j(x)$ by $x_o$, we obtain
\be
J_l \approx j(x_o) \left(\frac{-d^2 \ln j(x)}{dx^2}\Big|_{x_o} \right)^{-1}~,
\label{saddle}
\ee
where $x_o$ is a solution of the  equation, $dj/dx \big|_{x_o}=0$, 
\be
e^{{\tilde \beta} x_o} ({\tilde \beta} x_o-{\tilde \beta})+ \frac{\alpha^2}{{\tilde \beta}} ({\tilde \beta} x_o)^2 (1+x_o-{\tilde \beta} x_o)=0 ~.
\ee
As our numerics indicate that ${\tilde \beta} x_o$ is large and constant for a given $\alpha$, using the above equation, we obtain $ {\tilde \beta} x_o \approx 2 \ln \alpha$ for large $\alpha$. Using this solution in (\ref{saddle}), after some simplifications, we find that 
\be
J_l \propto x_o e^{-x_o}~.
\label{app_Jl}
\ee
Setting $J_l( {\tilde \beta} )=\rho$ in the above expression yields 
\be
{\tilde \beta} \approx \frac{\ln \alpha}{\ln \rho^{-1}}~.
\label{app_betastar}
\ee
Taking the derivative with respect to $\beta$ in (\ref{app_Jl}), we arrive at
\be
J_l'({\tilde \beta})= \frac{\rho (x_o-1)}{{\tilde \beta}} \sim \frac{\rho (\ln \rho)^2}{\ln \alpha}~.
\label{app_Jlprime}
\ee 
Our numerics are consistent with the above functional form and a numerical fit shows $J_l'({\tilde \beta})=(0.69/ \ln \alpha)~ \rho  ~[\ln(1.7/\rho)]^2$.


\clearpage

\begin{figure} 
\begin{center}  
\includegraphics[width=0.85\textwidth,angle=0,keepaspectratio=true]{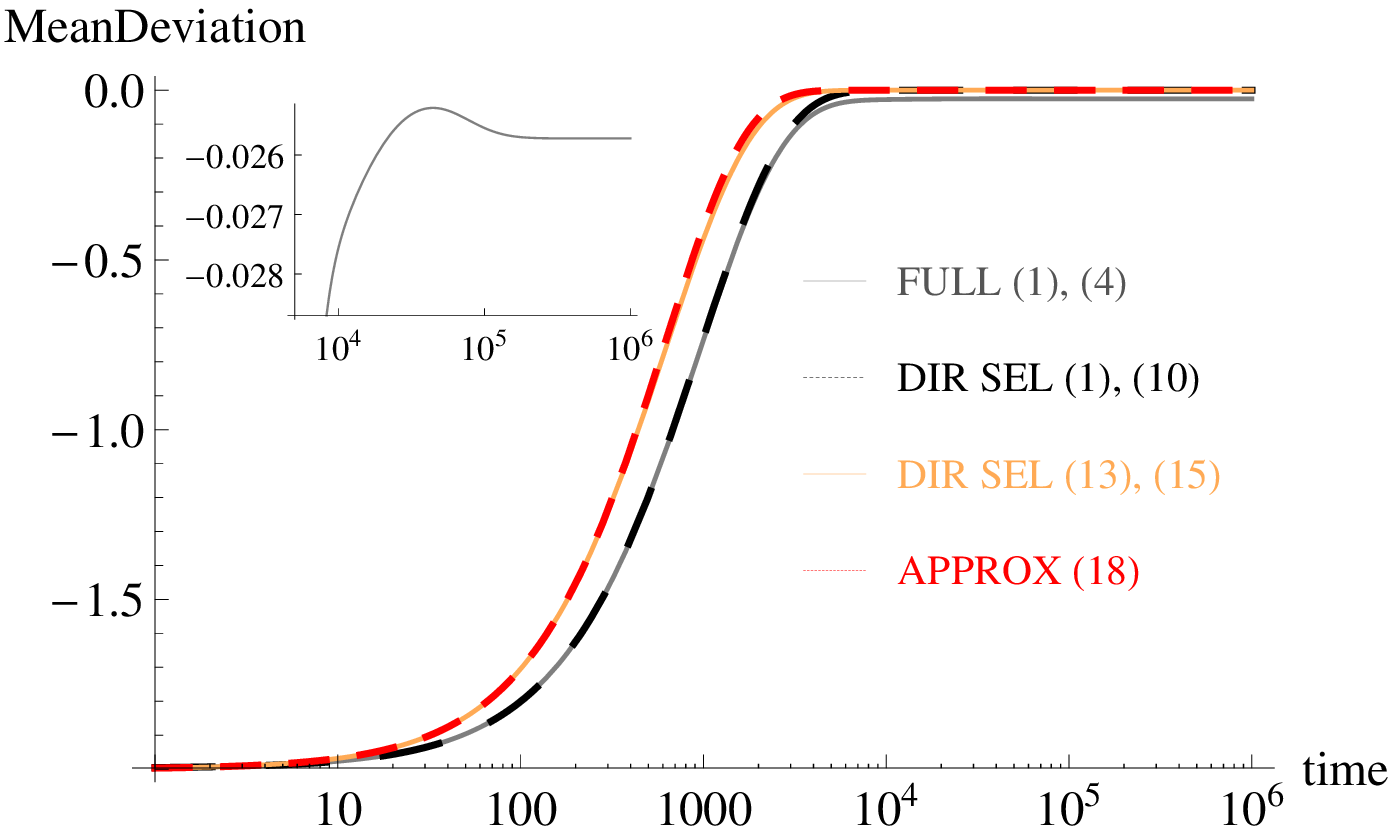}
\includegraphics[width=0.85\textwidth,angle=0,keepaspectratio=true]{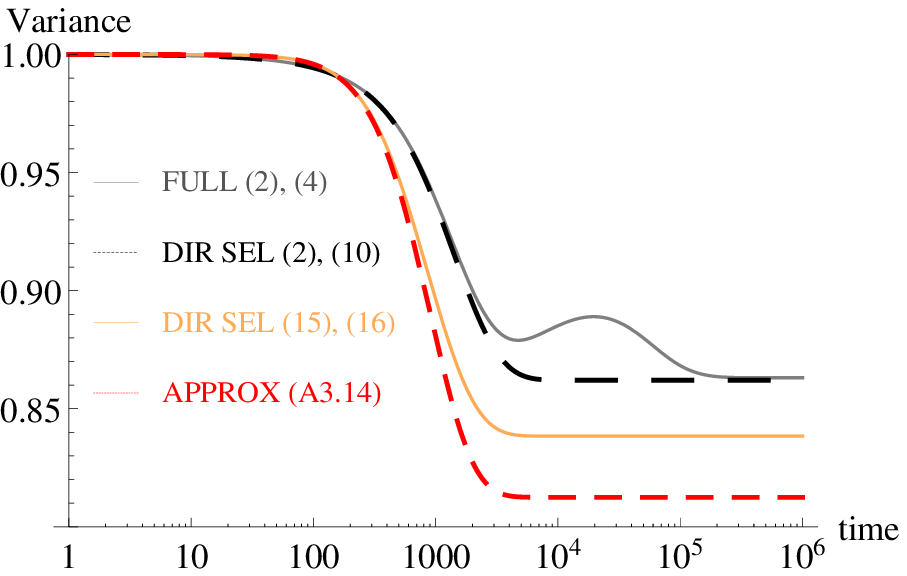}
 \end{center}
\caption{Most effects are small: Dynamics of the mean deviation and variance (relative to its  initial value) for ${\hat \gamma} \approx 3.2 {\bar \gamma} = 0.128$ and $z_f=\ell {\bar \gamma}/4=2$. The other parameters are $s=5 \times 10^{-3}, \mu=10^{-5}, {\bar \gamma}=0.04$. {The inset in the top panel shows the mean deviation in the full model at large times. The gray and black curves are obtained for a single  realization of effects in which $197$ out of $\ell=200$ loci have effects smaller than the mean. The orange and red curves, on the other hand, are averaged over the distribution of effects. The results from the two procedures match when the number of loci is very large as explained in DISCUSSION.}}
\label{fig_small_mean1}
\end{figure}
 
 \clearpage

\begin{figure} 
\begin{center}  
\includegraphics[width=1\textwidth,angle=0,keepaspectratio=true]{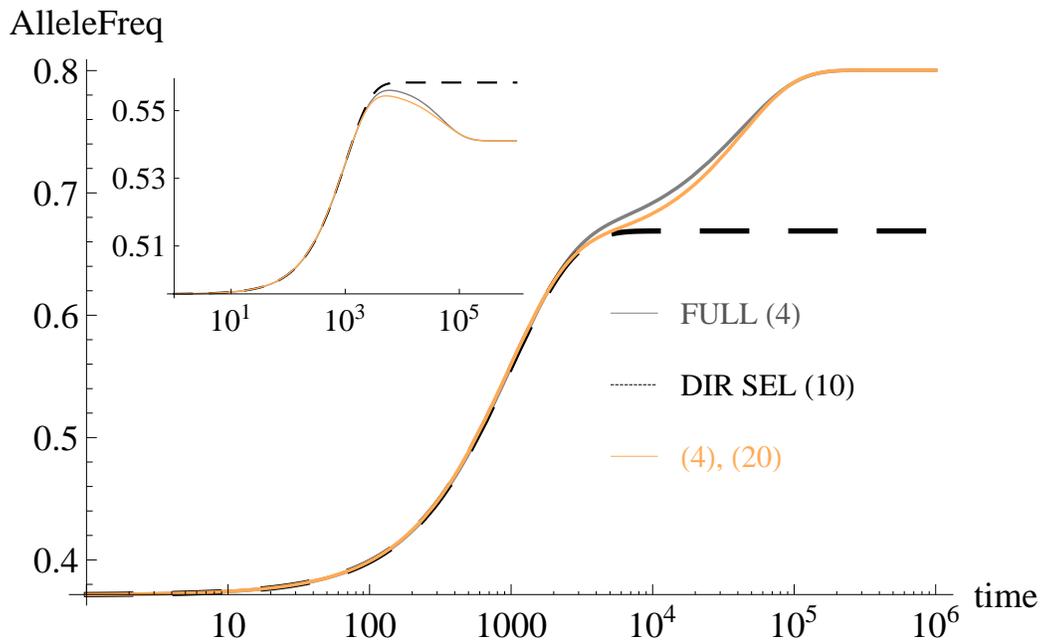}
\end{center}
\caption{Most effects are small: Dynamics of the allele frequency for locus with effect $\gamma_i=0.121$  (main) and $\gamma_i=0.024$ (inset). All the parameters are the same as in Fig.~\ref{fig_small_mean1}. {Note that while the initial allele frequency in the inset is closely approximated by (\ref{icss}), the allele trajectory starts at a value much lower than one half in the main panel as the effect size is close to ${\hat \gamma}\approx 0.126$. In the latter case, the corrections to (\ref{icss}) are substantial and given by (B2) of  \citet{Vladar:2014}.}}
\label{fig_small_freq}
\end{figure}

\clearpage

\begin{figure} 
\begin{center}  
\includegraphics[width=0.85\textwidth,angle=0,keepaspectratio=true]{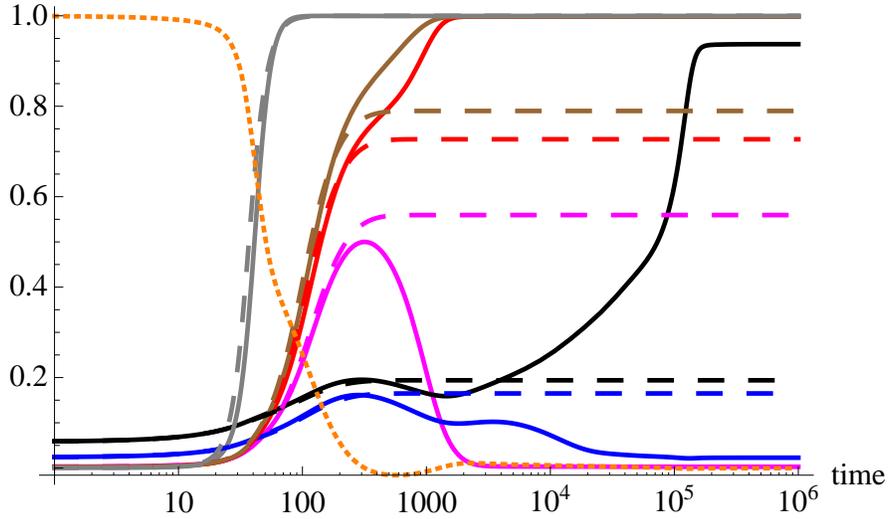}         
 \end{center}
\caption{Selective sweeps when most effects are large: Dynamics of  the scaled absolute mean deviation $|\Delta c_1(t)|/z_f$ (dotted, orange) and allele frequency (solid and dashed curves) for some loci that satisfy the necessary condition $\gamma_i < -2 \Delta c_1(0)$ where $\Delta c_1(0)=-2.99$ and $z_f=3$. The numerical solution of the full model (\ref{pevoleqnf}) (solid) and directional selection model (\ref{pevolshort}) (dashed) are shown for the effect size 
$\gamma_{i=1,...,6}=0.776$ (gray), $0.340$ (brown), $0.319$ (red), $0.272$ (magenta), $0.092$ (blue), $0.060$ (black). The other parameters are $s=0.1, \mu=10^{-4}, {\bar \gamma}=0.2, \ell=20$. Except for the locus with effect $\gamma_4$, the allele frequency for the directional selection model exceeds one half at loci whose allele frequency for the full model  sweeps. The allele frequency for the locus with effect $\gamma_6$ also increases nearly to fixation but at very long times where the directional selection model is not valid. Because this frequency increase is relatively slow, a genomic signature similar to a selective sweep cannot be expected.}
 \label{fig_large_freq1}
 \end{figure}

\clearpage

\begin{figure} 
\begin{center}  
\includegraphics[width=0.85\textwidth,angle=0,keepaspectratio=true]{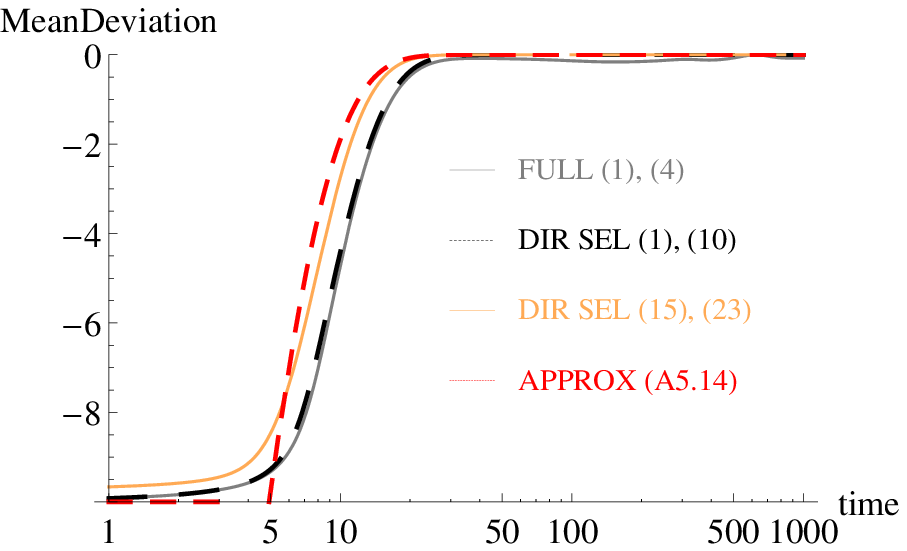}
\includegraphics[width=0.85\textwidth,angle=0,keepaspectratio=true]{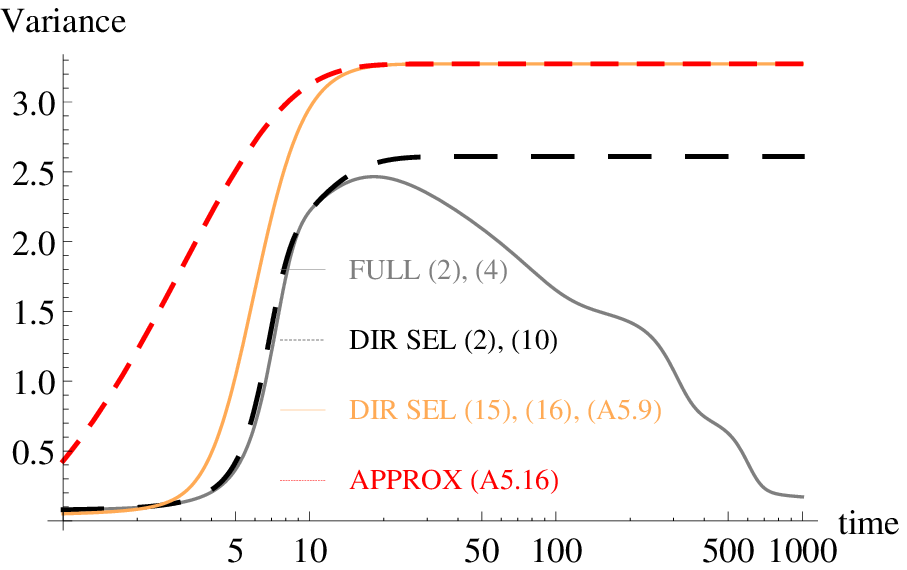}
\end{center}
\caption{Most effects are large: Dynamics of the mean deviation  for ${\hat \gamma} \approx 0.028, {\bar \gamma}=10 {\hat \gamma}$ and $z_f=10 \approx 0.18 \ell {\bar \gamma}$.  For the realization of effects used in this figure, $183$ out of $\ell=200$ loci have effects larger than the mean. The other parameters are $s=0.1, \mu=10^{-5}$ .}
\label{fig_small_meanL1}
\end{figure}

\clearpage

\begin{figure} 
\begin{center}  
\includegraphics[width=0.75\textwidth,angle=270,keepaspectratio=true]{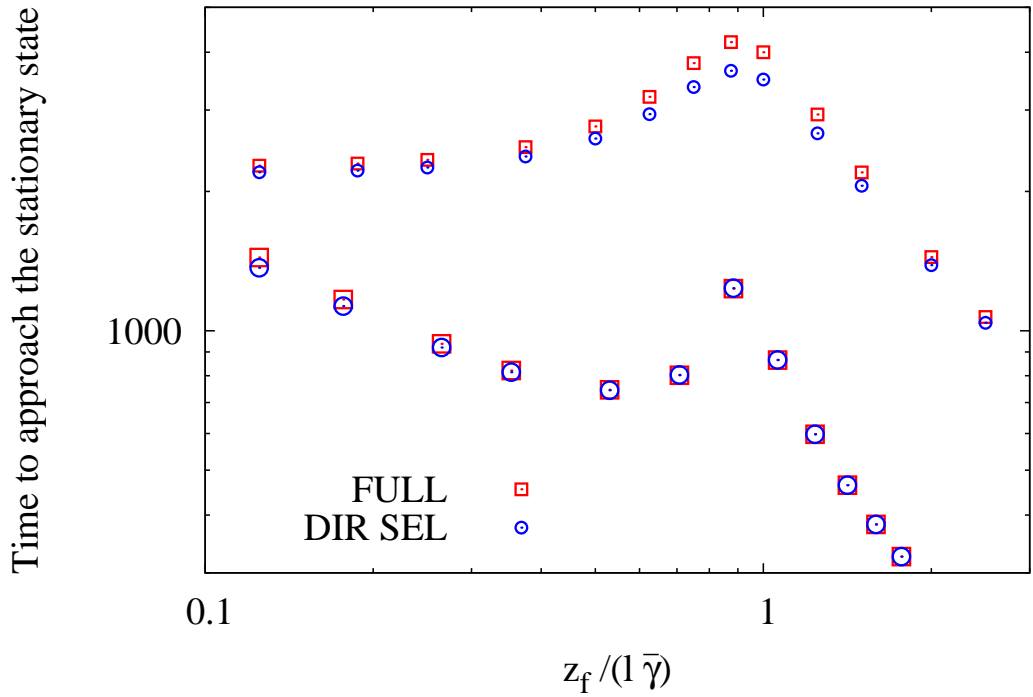}         
 \end{center}
\caption{Small vs. large effects when the number of loci are same: Time at which the mean reaches $0.9$ of the exact stationary state mean $c_1^*$ for various values of $z_f$ when most effects are small (small symbols) and large (large symbols). For the former (latter) case, the rest of the parameters are the same as in Fig.~\ref{fig_small_mean1} (\ref{fig_small_meanL1}). In order to show the data in the two cases on the same scale, the time for the large-effect case has been multiplied by $70$.}
\label{fig_zeffS}
\end{figure}

\clearpage

\begin{figure} 
\begin{center}  
\includegraphics[width=1\textwidth,angle=0,keepaspectratio=true]{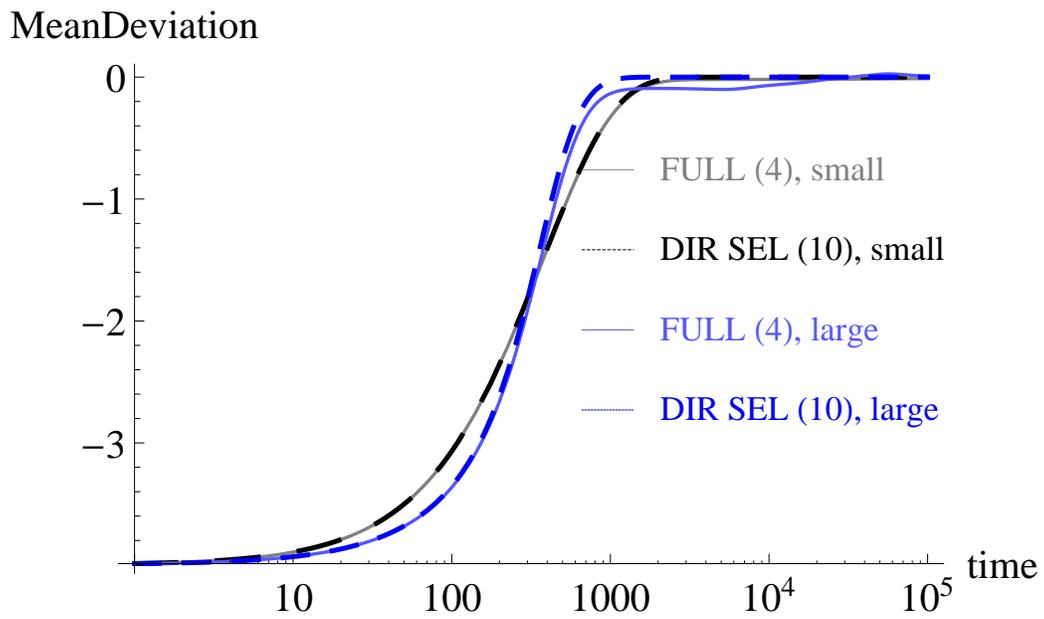}        
 \end{center}
\caption{Many loci with small effects vs. few loci with large effects: Mean deviation as a function of time when most effects are small (${\bar \gamma}=0.04, \ell=400$) and large (${\bar \gamma}=0.32, \ell=50$). In both cases, ${\hat \gamma} \approx 0.126$ ($s=5 \times 10^{-3}, \mu=10^{-5}$) and the new phenotypic optimum is shifted to $z_f= {\ell {\bar \gamma}}/4=4$.}
\label{fig_Fewl}
\end{figure}



\clearpage

\begin{thebibliography}{}

\bibitem[\protect\citeauthoryear{Abramowitz and Stegun}{\textsc{Abramowitz} and
  \textsc{Stegun}}{1964}]{Abramowitz:1964}
\textsc{Abramowitz, M.} and \textsc{I.~A. Stegun}, 1964\ \ {\em Handbook of
  Mathematical Functions with Formulas, Graphs, and Mathematical Tables}.
\newblock Dover.

\bibitem[\protect\citeauthoryear{Arfken}{\textsc{Arfken}}{1985}]{Arfken:1985}
\textsc{Arfken, G.}, 1985\ \ {\em Mathematical Methods for Physicists}.
\newblock Academic Press, New York.

\bibitem[\protect\citeauthoryear{Axelsson, Ratnakumar, Arendt, Maqbool,
  Webster, Perloski, Liberg, Arnemo, Hedhammar, and
  Lindblad-Toh}{\textsc{Axelsson} {\em et~al.\@}}{2013}]{Axelsson:2013}
\textsc{Axelsson, E.}, \textsc{A.~Ratnakumar}, \textsc{M.~L. Arendt},
  \textsc{K.~Maqbool}, \textsc{M.~T. Webster}, \textsc{M.~Perloski},
  \textsc{O.~Liberg}, \textsc{J.~M. Arnemo}, \textsc{A.~Hedhammar}, and
  \textsc{K.~Lindblad-Toh}, 2013\ \ The genomic signature of dog domestication
  reveals adaptation to a starch-rich diet.
\newblock Nature~{\bf 495}\textbf{:}\ 360--364.

\bibitem[\protect\citeauthoryear{Barton}{\textsc{Barton}}{1986}]{Barton:1986}
\textsc{Barton, N.~H.}, 1986\ \ The maintenance of polygenic variation through
  a balance between mutation and stabilizing selection.
\newblock Genet. Res.~{\bf 47}\textbf{:}\ 209--216.

\bibitem[\protect\citeauthoryear{Barton and de~Vladar}{\textsc{Barton} and
  \textsc{de~Vladar}}{2009}]{Barton:2009}
\textsc{Barton, N.~H.} and \textsc{H.~P. de~Vladar}, 2009\ \ Statistical
  Mechanics and the Evolution of Polygenic Quantitative Traits.
\newblock Genetics~{\bf 181}\textbf{:}\ 997--1011.

\bibitem[\protect\citeauthoryear{Barton and Turelli}{\textsc{Barton} and
  \textsc{Turelli}}{1991}]{Barton:1991}
\textsc{Barton, N.~H.} and \textsc{M.~Turelli}, 1991\ \ Natural and sexual
  selection on many loci.
\newblock Genetics~{\bf 127}\textbf{:}\ 229--255.

\bibitem[\protect\citeauthoryear{Bod{'}ov{\'a}, Tkacik, and
  Barton}{\textsc{Bod{'}ov{\'a}} {\em et~al.\@}}{2016}]{Budova:2016}
\textsc{Bod{'}ov{\'a}, K.}, \textsc{G.~Tkacik}, and \textsc{N.~H. Barton},
  2016\ \ A general approximation for the dynamics of quantitative traits.
\newblock Genetics~{\bf 202}\textbf{:}\ 1523--1548.

\bibitem[\protect\citeauthoryear{B{\"u}rger}{\textsc{B{\"u}rger}}{1991}]{Burger:1991}
\textsc{B{\"u}rger, R.}, 1991\ \ Moments, cumulants, and polygenic dynamics.
\newblock J. Math. Biol.~{\bf 30}\textbf{:}\ 199--213.

\bibitem[\protect\citeauthoryear{B{\"u}rger}{\textsc{B{\"u}rger}}{2000}]{Burger:2000}
\textsc{B{\"u}rger, R.}, 2000\ \ {\em The Mathematical Theory of Selection,
  Recombination, and Mutation}.
\newblock Wiley, Chichester.

\bibitem[\protect\citeauthoryear{Castellani and Cavagna}{\textsc{Castellani}
  and \textsc{Cavagna}}{2005}]{Castellani:2005}
\textsc{Castellani, T.} and \textsc{A.~Cavagna}, 2005\ \ Spin-glass theory for
  pedestrians.
\newblock J. Stat. Mech.:Theo. and Exp.~{\bf 2005}\textbf{:}\ P05012.

\bibitem[\protect\citeauthoryear{Charlesworth}{\textsc{Charlesworth}}{2013}]{Charlesworth:2013b}
\textsc{Charlesworth, B.}, 2013\ \ Stabilizing selection, purifying selection,
  and mutational bias in finite populations.
\newblock Genetics~{\bf 194}\textbf{:}\ 955--971.

\bibitem[\protect\citeauthoryear{Charlesworth and
  Charlesworth}{\textsc{Charlesworth} and
  \textsc{Charlesworth}}{2010}]{Charlesworth:2010}
\textsc{Charlesworth, B.} and \textsc{D.~Charlesworth}, 2010\ \ {\em Elements
  of evolutionary genetics}.
\newblock Roberts and Company Publishers.

\bibitem[\protect\citeauthoryear{Chevin and Hospital}{\textsc{Chevin} and
  \textsc{Hospital}}{2008}]{Chevin:2008}
\textsc{Chevin, L.-M.} and \textsc{F.~Hospital}, 2008\ \ Selective sweep at a
  quantitative trait locus in the presence of background genetic variation.
\newblock Genetics~{\bf 180}\textbf{:}\ 1645--1660.

\bibitem[\protect\citeauthoryear{Cook, Grant, Saccheri, and
  Mallet}{\textsc{Cook} {\em et~al.\@}}{2012}]{Cook:2012}
\textsc{Cook, L.~M.}, \textsc{B.~S. Grant}, \textsc{I.~J. Saccheri}, and
  \textsc{J.~Mallet}, 2012\ \ Selective bird predation on the peppered moth:
  the last experiment of {M}ichael {M}ajerus.
\newblock Biology Letters~{\bf 8}\textbf{:}\ 609--612.

\bibitem[\protect\citeauthoryear{Coop, Witonsky, Di~Rienzo, and
  Pritchard}{\textsc{Coop} {\em et~al.\@}}{2010}]{Coop:2010}
\textsc{Coop, G.}, \textsc{D.~Witonsky}, \textsc{A.~Di~Rienzo}, and
  \textsc{J.~K. Pritchard}, 2010\ \ Using environmental correlations to
  identify loci underlying local adaptation.
\newblock Genetics~{\bf 185}\textbf{:}\ 1411--1423.

\bibitem[\protect\citeauthoryear{de~Vladar and Barton}{\textsc{de~Vladar} and
  \textsc{Barton}}{2014}]{Vladar:2014}
\textsc{de~Vladar, H.~P.} and \textsc{N.~H. Barton}, 2014\ \ Stability and
  response of polygenic traits to stabilizing selection and mutation.
\newblock Genetics~{\bf 197}\textbf{:}\ 749--767.

\bibitem[\protect\citeauthoryear{Ffrench-Constant, Bogwitz, Daborne, and
  Yen}{\textsc{Ffrench-Constant} {\em et~al.\@}}{2002}]{Ffrench:2002}
\textsc{Ffrench-Constant, R.~H.}, \textsc{M.~Bogwitz}, \textsc{P.~Daborne}, and
  \textsc{J.~Yen}, 2002\ \ A single {P}450 allele associated with insecticide
  resistance in {D}rosophila.
\newblock Science~{\bf 27}\textbf{:}\ 2253--2256.

\bibitem[\protect\citeauthoryear{Foll and Gaggiotti}{\textsc{Foll} and
  \textsc{Gaggiotti}}{2008}]{Foll:2008}
\textsc{Foll, M.} and \textsc{O.~Gaggiotti}, 2008\ \ A genome-scan method to
  identify selected loci appropriate for both dominant and codominant markers:
  a {B}ayesian perspective.
\newblock Genetics~{\bf 180}\textbf{:}\ 977--993.

\bibitem[\protect\citeauthoryear{Goddard and Hayes}{\textsc{Goddard} and
  \textsc{Hayes}}{2009}]{Goddard:2009}
\textsc{Goddard, M.~E.} and \textsc{B.~J. Hayes}, 2009\ \ Mapping genes for
  complex traits in domestic animals and their use in breeding programmes.
\newblock Nat. Rev. Genet.~{\bf 10}\textbf{:}\ 381--391.

\bibitem[\protect\citeauthoryear{Gomulkiewicz, Holt, Barfield, and
  Nuismer}{\textsc{Gomulkiewicz} {\em et~al.\@}}{2010}]{Gomulkiewicz:2010}
\textsc{Gomulkiewicz, R.}, \textsc{R.~D. Holt}, \textsc{M.~Barfield}, and
  \textsc{S.~L. Nuismer}, 2010\ \ Genetics, adaptation, and invasion in harsh
  environments.
\newblock Evolutionary Applications~{\bf 3}\textbf{:}\ 97--108.

\bibitem[\protect\citeauthoryear{Grant and Grant}{\textsc{Grant} and
  \textsc{Grant}}{2008}]{Grant:2008}
\textsc{Grant, P.~R.} and \textsc{B.~R. Grant}, 2008\ \ {\em How and Why
  Species Multiply: The Radiation of DarwinÕs Finches}.
\newblock Princeton NJ: Princeton University Press.

\bibitem[\protect\citeauthoryear{Jain and Stephan}{\textsc{Jain} and
  \textsc{Stephan}}{2015}]{Jain:2015}
\textsc{Jain, K.} and \textsc{W.~Stephan}, 2015\ \ Response of polygenic traits
  under stabilising selection and mutation when loci have unequal effects.
\newblock G3: Genes, Genomes, Genetics~{\bf 5}\textbf{:}\ 1065--1074.

\bibitem[\protect\citeauthoryear{Kaplan, Hudson, and Langley}{\textsc{Kaplan}
  {\em et~al.\@}}{1989}]{Kaplan:1989}
\textsc{Kaplan, N.~L.}, \textsc{R.~R. Hudson}, and \textsc{C.~H. Langley},
  1989\ \ The {``}hitchhiking effect{"} revisited.
\newblock Genetics~{\bf 123}\textbf{:}\ 887--899.

\bibitem[\protect\citeauthoryear{Kim and Stephan}{\textsc{Kim} and
  \textsc{Stephan}}{2002}]{Kim:2002}
\textsc{Kim, Y.} and \textsc{W.~Stephan}, 2002\ \ Detecting a local signature
  of genetic hitchhiking along a recombining chromosome.
\newblock Genetics~{\bf 160}\textbf{:}\ 765--777.

\bibitem[\protect\citeauthoryear{Lamichhaney, Berglund, Alm{\'e}n, Maqbool,
  Grabherr, Martinez-Barrio, Promerov{\'a}, Rubin, Wang, Zamani, Grant, Grant,
  Webster, and Andersson}{\textsc{Lamichhaney} {\em
  et~al.\@}}{2015}]{Lamichhaney:2015}
\textsc{Lamichhaney, S.}, \textsc{J.~Berglund}, \textsc{M.~S. Alm{\'e}n},
  \textsc{K.~Maqbool}, \textsc{M.~Grabherr}, \textsc{A.~Martinez-Barrio},
  \textsc{M.~Promerov{\'a}}, \textsc{C.~J. Rubin}, \textsc{C.~Wang},
  \textsc{N.~Zamani}, \textsc{B.~R. Grant}, \textsc{P.~R. Grant}, \textsc{M.~T.
  Webster}, and \textsc{L.~Andersson}, 2015\ \ Evolution of {D}arwin's finches
  and their beaks revealed by genome sequencing.
\newblock Nature~{\bf 518}\textbf{:}\ 371--375.

\bibitem[\protect\citeauthoryear{Lamichhaney, Martinez~Barrio, Rafati,
  Sundstr{\"o}m, Rubin, Gilbert, Berglund, Wetterbom, Laikre, Webster,
  Grabherr, Ryman, and Andersson}{\textsc{Lamichhaney} {\em
  et~al.\@}}{2012}]{Lamichhaney:2012}
\textsc{Lamichhaney, S.}, \textsc{A.~Martinez~Barrio}, \textsc{N.~Rafati},
  \textsc{G.~Sundstr{\"o}m}, \textsc{C.~J. Rubin}, \textsc{E.~R. Gilbert},
  \textsc{J.~Berglund}, \textsc{A.~Wetterbom}, \textsc{L.~Laikre},
  \textsc{M.~T. Webster}, \textsc{M.~Grabherr}, \textsc{N.~Ryman}, and
  \textsc{L.~Andersson}, 2012\ \ Population-scale sequencing reveals genetic
  differentiation due to local adaptation in {A}tlantic herring.
\newblock Proc. Natl. Acad. Sci. USA~{\bf 109}\textbf{:}\ 19345--19350.

\bibitem[\protect\citeauthoryear{Lande}{\textsc{Lande}}{1976}]{Lande:1976}
\textsc{Lande, R.}, 1976\ \ Natural selection and random genetic drift in
  phenotypic evolution.
\newblock Evolution~{\bf 30}\textbf{:}\ 314--334.

\bibitem[\protect\citeauthoryear{Lande}{\textsc{Lande}}{1983}]{Lande:1983}
\textsc{Lande, R.}, 1983\ \ The response to selection on major and minor
  mutations affecting a metrical trait.
\newblock Heredity~{\bf 50}\textbf{:}\ 47--65.

\bibitem[\protect\citeauthoryear{Linnen, Poh, Peterson, Barrett, Larson,
  Jensen, and Hoekstra}{\textsc{Linnen} {\em et~al.\@}}{2013}]{Linnen:2013}
\textsc{Linnen, C.~R.}, \textsc{Y.~P. Poh}, \textsc{B.~K. Peterson},
  \textsc{R.~D. Barrett}, \textsc{J.~G. Larson}, \textsc{J.~D. Jensen}, and
  \textsc{H.~E. Hoekstra}, 2013\ \ Adaptive evolution of multiple traits
  through multiple mutations at a single gene.
\newblock Science~{\bf 339}\textbf{:}\ 1312--1316.

\bibitem[\protect\citeauthoryear{Losos}{\textsc{Losos}}{2009}]{Losos:2009}
\textsc{Losos, J.~B.}, 2009\ \ {\em Lizards in an Evolutionary Tree: Ecology
  and Adaptive Radiation of Anoles}.
\newblock San Francisco, CA: University of California Press.

\bibitem[\protect\citeauthoryear{Losos}{\textsc{Losos}}{2014}]{Losos:2014}
\textsc{Losos, J.~B.}, 2014\ \ What Darwin got wrong.
\newblock Chronicle of Higher Education~{\bf -}\textbf{:}\ B13--B15.

\bibitem[\protect\citeauthoryear{Mackay}{\textsc{Mackay}}{2004}]{Mackay:2004}
\textsc{Mackay, T. F.~C.}, 2004\ \ The genetic architecture of quantitative
  traits: lessons from {\it Drosophila}.
\newblock Current Opinion in Genetics and Development~{\bf 14}\textbf{:}\
  253--257.

\bibitem[\protect\citeauthoryear{Matuszewski, Hermisson, and
  Kopp}{\textsc{Matuszewski} {\em et~al.\@}}{2015}]{Matuszewski:2015}
\textsc{Matuszewski, S.}, \textsc{J.~Hermisson}, and \textsc{M.~Kopp}, 2015\ \
  Catch me if you can: adaptation from standing genetic variation to a moving
  phenotypic optimum.
\newblock Genetics~{\bf 200}\textbf{:}\ 1255--1274.

\bibitem[\protect\citeauthoryear{Maynard~Smith and
  Haigh}{\textsc{Maynard~Smith} and \textsc{Haigh}}{1974}]{Smith:1974}
\textsc{Maynard~Smith, J.} and \textsc{J.~Haigh}, 1974\ \ Hitchhiking effect of
  a favourable gene.
\newblock Genet. Res.~{\bf 23}\textbf{:}\ 23--35.

\bibitem[\protect\citeauthoryear{Nielsen, Williamson, Kim, Hubisz, Clark, and
  Bustamante}{\textsc{Nielsen} {\em et~al.\@}}{2005}]{Nielsen:2005b}
\textsc{Nielsen, R.}, \textsc{S.~Williamson}, \textsc{Y.~Kim}, \textsc{M.~J.
  Hubisz}, \textsc{A.~G. Clark}, and \textsc{C.~Bustamante}, 2005\ \ Genomic
  scans for selective sweeps using {SNP} data.
\newblock Genome Res.~{\bf 15}\textbf{:}\ 1566--1575.

\bibitem[\protect\citeauthoryear{Pavlidis, Jensen, and
  Stephan}{\textsc{Pavlidis} {\em et~al.\@}}{2010}]{Pavlidis:2010}
\textsc{Pavlidis, P.}, \textsc{J.~D. Jensen}, and \textsc{W.~Stephan}, 2010\ \
  Searching for footprints of positive selection in whole-genome SNP data from
  nonequilibrium populations.
\newblock Genetics~{\bf 185}\textbf{:}\ 907--922.

\bibitem[\protect\citeauthoryear{Pavlidis, Metzler, and
  Stephan}{\textsc{Pavlidis} {\em et~al.\@}}{2012}]{Pavlidis:2012}
\textsc{Pavlidis, P.}, \textsc{D.~Metzler}, and \textsc{W.~Stephan}, 2012\ \
  Selective sweeps in multilocus models of quantitative traits.
\newblock Genetics~{\bf 192}\textbf{:}\ 225--239.

\bibitem[\protect\citeauthoryear{Pritchard and Di~Rienzo}{\textsc{Pritchard}
  and \textsc{Di~Rienzo}}{2010}]{Pritchard:2010}
\textsc{Pritchard, J.~K.} and \textsc{A.~Di~Rienzo}, 2010\ \ Adaptation {-} not
  by sweeps alone.
\newblock Nature Review Genetics~{\bf 11}\textbf{:}\ 665--667.

\bibitem[\protect\citeauthoryear{Reznick}{\textsc{Reznick}}{2009}]{Reznick:2009}
\textsc{Reznick, D.~N.}, 2009\ \ {\em The Origin Then and Now: An Interpretive
  Guide to the Origin of Species}.
\newblock Princeton NJ: Princeton University Press.

\bibitem[\protect\citeauthoryear{Riebler, Held, and Stephan}{\textsc{Riebler}
  {\em et~al.\@}}{2008}]{Riebler:2008}
\textsc{Riebler, A.}, \textsc{L.~Held}, and \textsc{W.~Stephan}, 2008\ \
  {B}ayesian variable selection for detecting adaptive genomic differences
  among populations.
\newblock Genetics~{\bf 178}\textbf{:}\ 1817--1829.

\bibitem[\protect\citeauthoryear{Rubin, Megens, Martinez~Barrio, Maqbool,
  Sayyab, Schwochow, Wang, Carlborg, Jern, J{\o}rgensen, Archibald, Fredholm,
  Groenen, and Andersson}{\textsc{Rubin} {\em et~al.\@}}{2012}]{Rubin:2012}
\textsc{Rubin, C.~J.}, \textsc{H.~J. Megens}, \textsc{A.~Martinez~Barrio},
  \textsc{K.~Maqbool}, \textsc{S.~Sayyab}, \textsc{D.~Schwochow},
  \textsc{C.~Wang}, \textsc{{\"O}.~Carlborg}, \textsc{P.~Jern}, \textsc{C.~B.
  J{\o}rgensen}, \textsc{A.~L. Archibald}, \textsc{M.~Fredholm}, \textsc{M.~A.
  Groenen}, and \textsc{L.~Andersson}, 2012\ \ Strong signatures of selection
  in the domestic pig genome.
\newblock Proc. Natl. Acad. Sci. USA~{\bf 109}\textbf{:}\ 19529--19536.

\bibitem[\protect\citeauthoryear{Sornette}{\textsc{Sornette}}{2000}]{Sornette:2000}
\textsc{Sornette, D.}, 2000\ \ {\em Critical Phenomena in Natural Sciences}.
\newblock Springer, Berlin.

\bibitem[\protect\citeauthoryear{Stephan, Wiehe, and Lenz}{\textsc{Stephan}
  {\em et~al.\@}}{1992}]{Stephan:1992}
\textsc{Stephan, W.}, \textsc{T.~H.~E. Wiehe}, and \textsc{M.~W. Lenz}, 1992\ \
  The effect of strongly selected substitutions on neutral polymorphism:
  analytical results based on diffusion theory.
\newblock Theor. Pop. Biol.~{\bf 41}\textbf{:}\ 237--254.

\bibitem[\protect\citeauthoryear{Svetec, Werzner, Wilches, Pavlidis,
  Alvarez-Castro, Broman, Metzler, and Stephan}{\textsc{Svetec} {\em
  et~al.\@}}{2011}]{Svetec:2011}
\textsc{Svetec, N.}, \textsc{A.~Werzner}, \textsc{R.~Wilches},
  \textsc{P.~Pavlidis}, \textsc{J.~M. Alvarez-Castro}, \textsc{K.~W. Broman},
  \textsc{D.~Metzler}, and \textsc{W.~Stephan}, 2011\ \ Identification of
  {X}-linked quantitative trait loci affecting cold tolerance in {D}rosophila
  melanogaster and fine mapping by selective sweep analysis.
\newblock Mol Ecol.~{\bf 20}\textbf{:}\ 530--544.

\bibitem[\protect\citeauthoryear{Turchin, Chiang, Palmer, Sankararaman, Reich,
  and Hirschhorn}{\textsc{Turchin} {\em et~al.\@}}{2012}]{Turchin:2012}
\textsc{Turchin, M.~C.}, \textsc{C.~W.~K. Chiang}, \textsc{C.~D. Palmer},
  \textsc{S.~Sankararaman}, \textsc{D.~Reich}, and \textsc{J.~Hirschhorn},
  2012\ \ Evidence of widespread selection on standing variation in {E}urope at
  height-associated {SNP}s.
\newblock Nat. Genet.~{\bf 44}\textbf{:}\ 1015--1019.

\bibitem[\protect\citeauthoryear{van't Hof, Edmonds, Dalikov{\'a}, Marec, and
  Saccheri}{\textsc{van't Hof} {\em et~al.\@}}{2011}]{Hof:2011}
\textsc{van't Hof, A.~E.}, \textsc{N.~Edmonds}, \textsc{M.~Dalikov{\'a}},
  \textsc{F.~Marec}, and \textsc{I.~J. Saccheri}, 2011\ \ Industrial melanism
  in {B}ritish peppered moths has a singular and recent mutational origin.
\newblock Science~{\bf 332}\textbf{:}\ 958--960.

\bibitem[\protect\citeauthoryear{Vignieri, Larson, and
  Hoekstra}{\textsc{Vignieri} {\em et~al.\@}}{2010}]{Vignieri:2010}
\textsc{Vignieri, S.~N.}, \textsc{J.~G. Larson}, and \textsc{H.~E. Hoekstra},
  2010\ \ The selective advantage of crypsis in mice.
\newblock Evolution~{\bf 64}\textbf{:}\ 2153--2158.

\bibitem[\protect\citeauthoryear{Visscher, Brown, McCarthy, and
  Yang}{\textsc{Visscher} {\em et~al.\@}}{2012}]{Visscher:2012}
\textsc{Visscher, P.~M.}, \textsc{M.~A. Brown}, \textsc{M.~I. McCarthy}, and
  \textsc{J.~Yang}, 2012\ \ Five years of {GWAS} discovery.
\newblock Am J Hum Genet.~{\bf 90}\textbf{:}\ 7--24.

\bibitem[\protect\citeauthoryear{Wilches, Voigt, Duchen, Laurent, and
  Stephan}{\textsc{Wilches} {\em et~al.\@}}{2014}]{Wilches:2014}
\textsc{Wilches, R.}, \textsc{S.~Voigt}, \textsc{P.~Duchen},
  \textsc{S.~Laurent}, and \textsc{W.~Stephan}, 2014\ \ Fine-mapping and
  selective sweep analysis of {QTL} for cold tolerance in {D}rososphila
  melanogaster.
\newblock G3: Genes, Genomes, Genetics~{\bf 4}\textbf{:}\ 1635--1645.

\bibitem[\protect\citeauthoryear{Wollstein and Stephan}{\textsc{Wollstein} and
  \textsc{Stephan}}{2014}]{Wollstein:2014}
\textsc{Wollstein, A.} and \textsc{W.~Stephan}, 2014\ \ Adaptive fixation in
  two-locus models of stabilizing selection and genetic drift.
\newblock Genetics~{\bf 198}\textbf{:}\ 685--697.

\bibitem[\protect\citeauthoryear{Wright}{\textsc{Wright}}{1935}]{Wright:1935}
\textsc{Wright, S.}, 1935\ \ Evolution in populations in approximate
  equilibrium.
\newblock J. Genet.~{\bf 30}\textbf{:}\ 257--266.

\end{thebibliography}


\clearpage

\begin{center}

{\huge{\bf File S1}}

\hspace{7in}

{\huge Rapid adaptation of a polygenic trait after a sudden environmental shift}

\hspace{1in}

{\huge Supporting Information} 

\hspace{2in}

{\large Kavita Jain and Wolfgang Stephan}
\end{center}

\setcounter{figure}{0}
\setcounter{section}{0}

\setcounter{page}{1}
\makeatletter 
\renewcommand{\thefigure}{S\@arabic\c@figure} 
\renewcommand{\thesection}{S\@arabic\c@section} 
\renewcommand{\thetable}{S\@arabic\c@table} 

\rfoot{\thepage}
\lfoot{K. Jain and W. Stephan}

\begin{figure} 
\begin{center}  
\includegraphics[width=0.75\textwidth,angle=270,keepaspectratio=true]{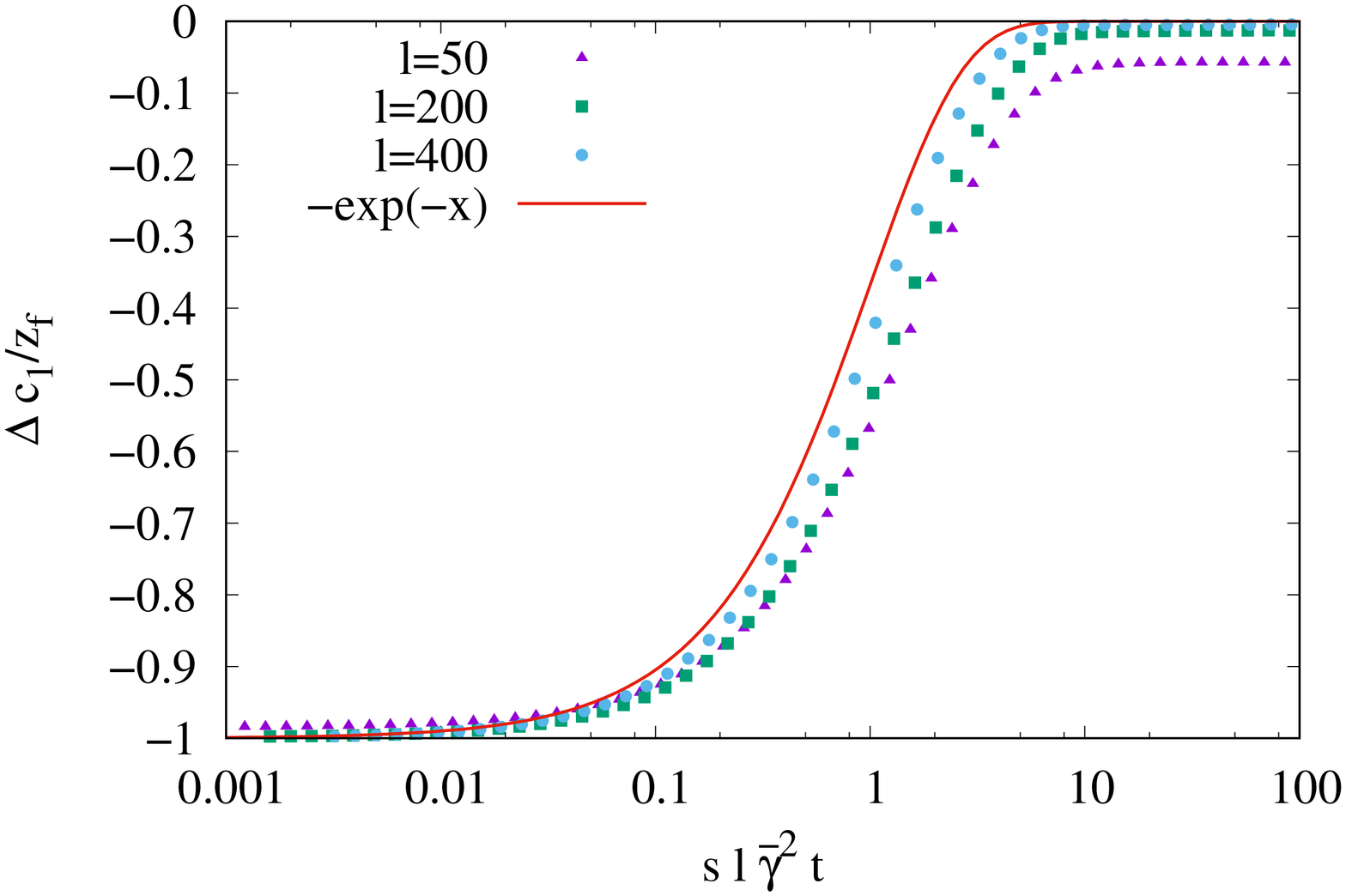}         
 \end{center}
\caption{Mean deviation for various values of the number of loci and fixed $z_f={\ell {\bar \gamma}}/4$  for the parameters in Fig.~\ref{fig_small_mean1} to show that the numerical data obtained in the full model (\ref{pevoleqnf}) for a single set of effects approaches the analytical expression (\ref{small_mean}) with increasing $\ell$. }
 \label{fig_infL}
 \end{figure}

 \clearpage
 
 \section{Cumulant hierarchy}
\label{app_cumhie}

Here we briefly discuss a commonly used method and its inadequacy. 
The equation  (\ref{cumshort}) for the $n$th cumulant is related to one higher one and to break this hierarchy, one may set all the cumulants higher than $n^*$ to be zero. When most effects are small, the dynamics can be described by setting the second cumulant to be constant and all the higher cumulants to be zero ($n^*=2$). However, when most effects are large, the standard approximation of terminating the infinite set of differential equations for the cumulants by setting $c_n(t)=0, n > n^*$ can not capture these dynamics. 
As Fig.~\ref{fig_closure3} shows, the essential features of the cumulants are not captured even when $n^*=9$. 

This rather unexpected result can be understood by noting that the cumulants (except mean) vary {\it nonmonotonically} with time as  shown in Figure~\ref{fig_closure3}. Around $t=50$, the variance has a maximum but the mean has not equilibrated. Then by (\ref{cumshort}) for $n=2$, the LHS is zero at $t=50$ and therefore the third cumulant must vanish as indeed seen in Fig.~\ref{fig_closure3}. But if we were to choose $n^*=3$ (i.e., $c_3(t)=c_3(0), c_n(t)=0, n > 3$), the derivative of the variance can never be zero and the nonmonotonic behavior of the variance can not be captured by such an approximation. A similar argument holds for higher cumulants since they also oscillate in time. 

 \clearpage

\begin{figure} 
\begin{center}  
\includegraphics[width=0.75\textwidth,angle=0,keepaspectratio=true]{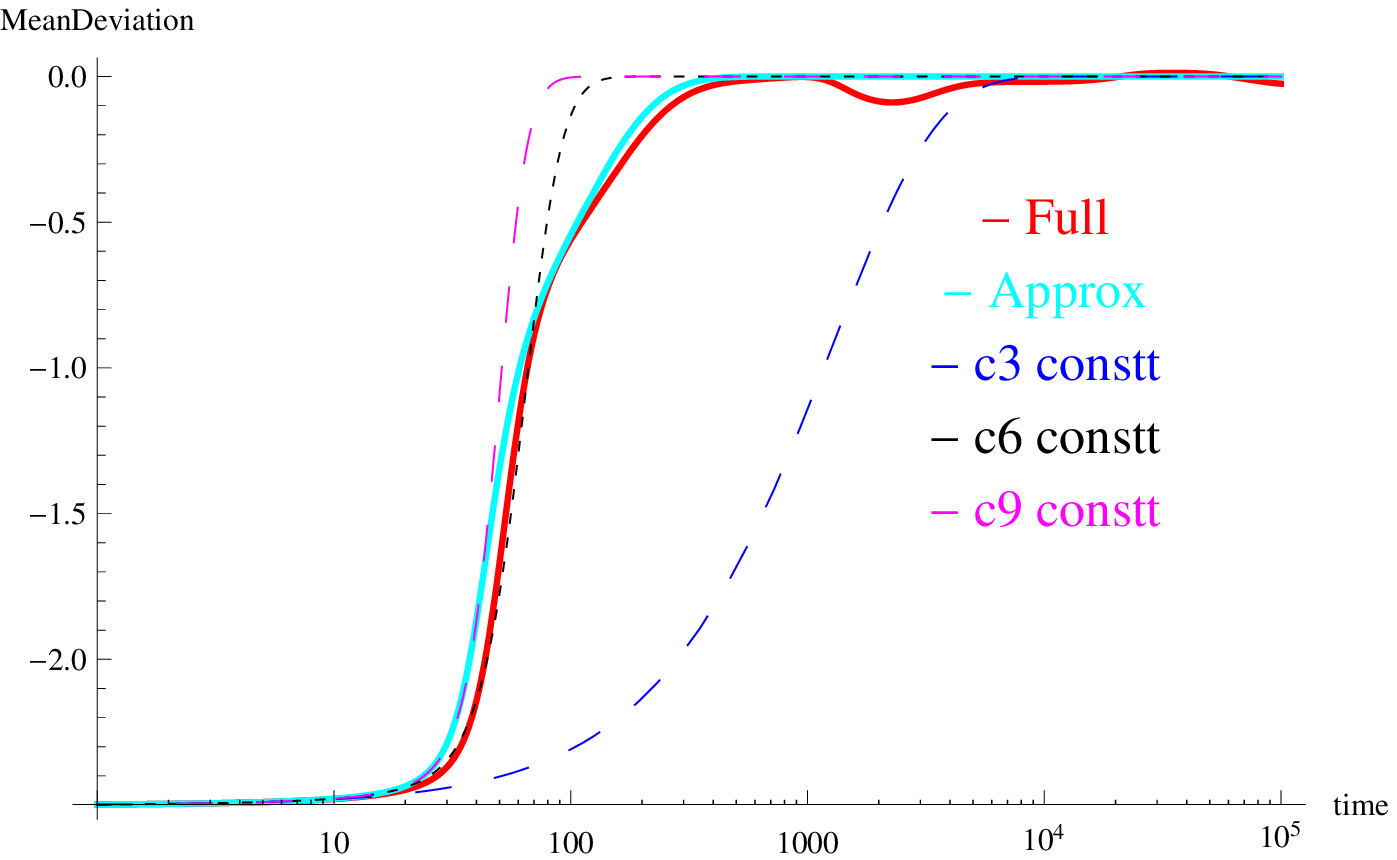}         
\includegraphics[width=0.75\textwidth,angle=0,keepaspectratio=true]{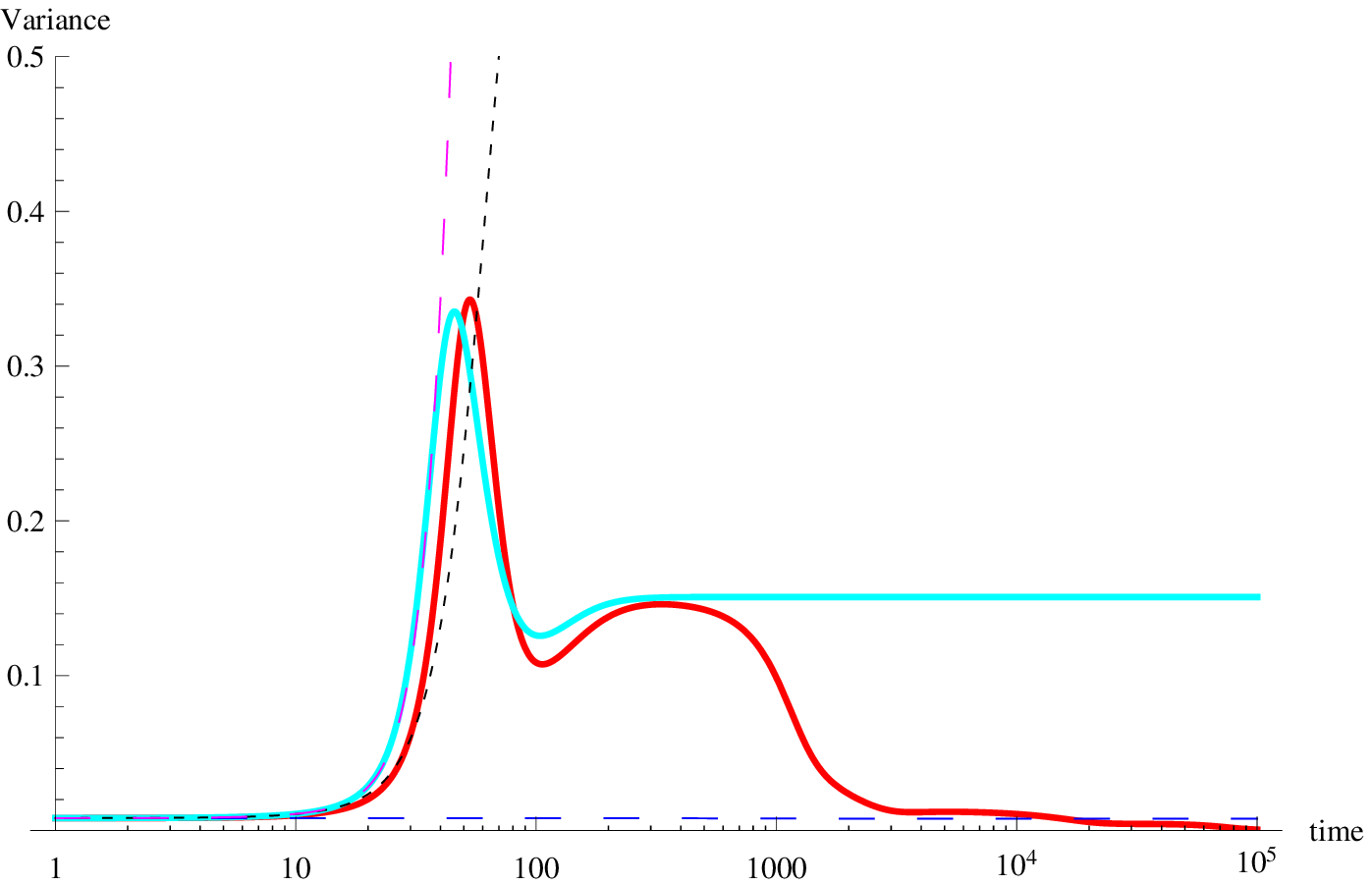} 
\includegraphics[width=0.75\textwidth,angle=0,keepaspectratio=true]{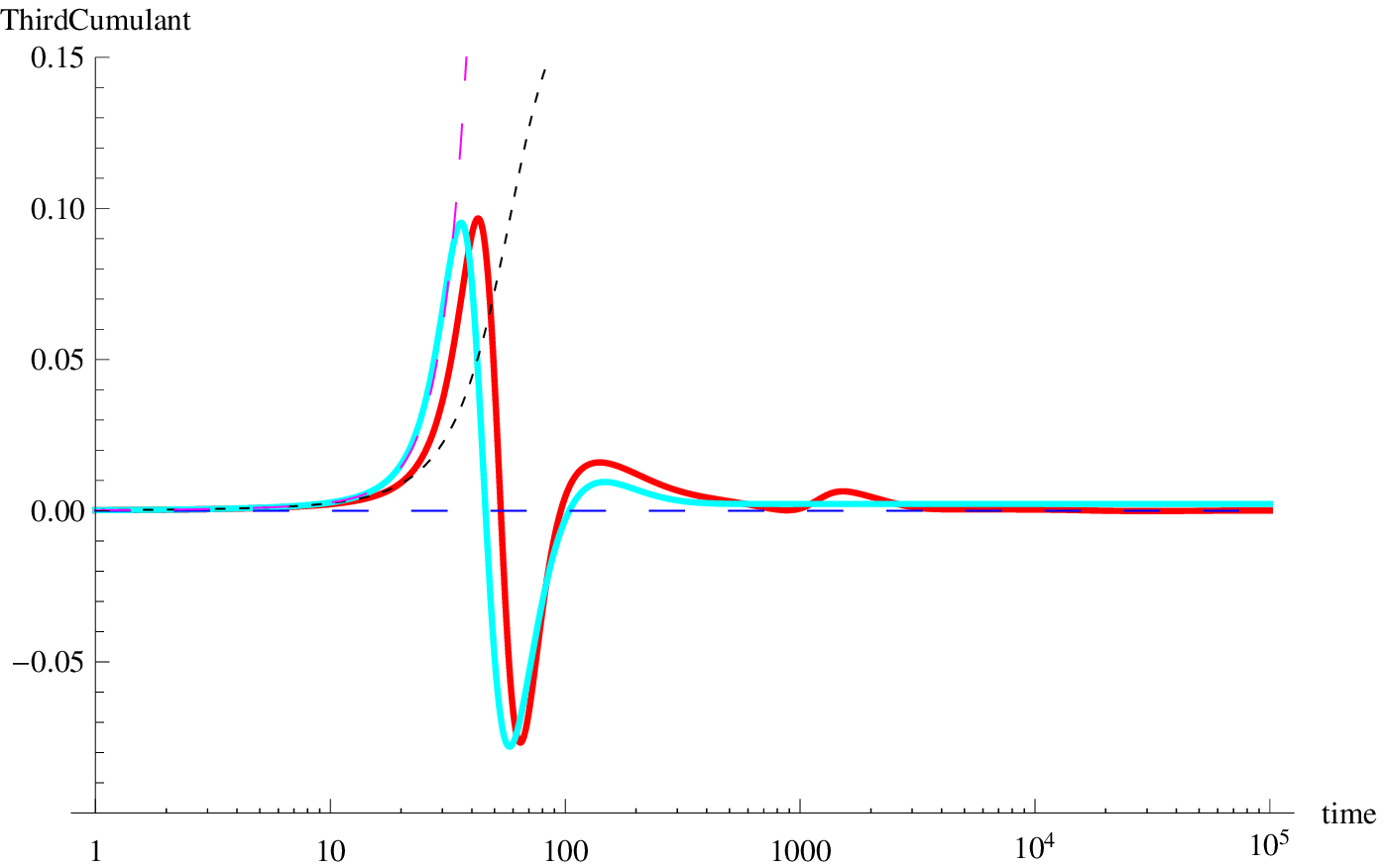} 
 \end{center}
\caption{Dynamics of the cumulants when most effects are large: Numerical results obtained by closing the hierarchy in (\ref{cumshort}) when the third, sixth and ninth cumulants are assumed to be constant. Full and Approx refer to results obtained using the full model (\ref{pevoleqnf}) and directional selection model (\ref{pevolshort}). The parameters are the same as in Fig.~\ref{fig_large_freq1}.}
 \label{fig_closure3}
 \end{figure}

%
 
\clearpage

 \begin{table}[t]
\begin{center}
\begin{tabular}{|c|c|c|}
\hline
$\alpha$ & $\beta$   & Integral $J_l$  \\
\hline
\hline
20 & 0.1 & 0.00646802 \\
& 0.5 & 0.00782825 \\
& 1 &  0.0132363\\
& 2&0.108583 \\
& 3&0.313421 \\
&4&0.506622\\
&5&0.650011\\
&6&0.74967\\
\hline
200 & 0.1 &  0.000120827\\
& 0.5 & 0.000135405  \\
& 1 & 0.000201675 \\
& 2&0.00935999 \\
&3&0.0704926 \\
&4&0.181047\\
&5&0.304219\\
&6&0.417745\\
\hline
2000 & 0.1 & $1.78243 \times 10^{-6}$ \\
& 0.5 &  $1.92926 \times 10^{-6}$\\
& 1 & $2.6638 \times 10^{-6}$ \\
& 2&0.000880179 \\
&3&0.0160675\\
&4&0.0635073\\
&5&0.137642\\
&6&0.223307\\
\hline
\hline
\end{tabular}
\caption{Numerical evaluation of the integral $J_l$ defined by (\ref{Jldefn}) for various $\alpha$ and $\beta$.}
\label{tab_Jl}
\end{center}
\end{table}

\end{document}